\documentclass[apj]{emulateapj}
\usepackage[]{units}
\usepackage[update,prepend]{epstopdf}
\usepackage{graphicx}
\usepackage{amssymb}
\usepackage{amsmath}
\usepackage{threeparttable}
\usepackage{hyperref}
\usepackage{rotating}
\usepackage{color}
\usepackage{pifont}
\newcommand{\cmark}{\ding{51}}%
\newcommand{\xmark}{\ding{55}}%

\def\aap{A\&A}

\def\apjl{ApJ}

\def\apjs{ApJS}

\shorttitle{X-rays of extremely red quasars}
\shortauthors{Goulding et al.}

\begin{document}

\title{High redshift extremely red quasars in X-rays}

\author{Andy D. Goulding\altaffilmark{1}, Nadia L. Zakamska\altaffilmark{2,3}, Rachael M. Alexandroff\altaffilmark{4,2}, Roberto J. Assef\altaffilmark{5}, Manda Banerji\altaffilmark{6}, Fred Hamann\altaffilmark{7}, Dominika Wylezalek\altaffilmark{2,10}, William N. Brandt\altaffilmark{8}, Jenny E. Greene\altaffilmark{1}, George B. Lansbury\altaffilmark{6,9}, Isabelle P\^aris\altaffilmark{11}, Gordon Richards\altaffilmark{12}, Daniel Stern\altaffilmark{13}, Michael A. Strauss\altaffilmark{1}}

\affil{$^1$Department of Astrophysical Sciences, Princeton University, Princeton, NJ 08540, USA; $^{\dagger}$goulding@astro.princeton.edu}
\affil{$^2$Department of Physics \& Astronomy, Johns Hopkins University, Bloomberg Center, 3400 N. Charles St., Baltimore, MD 21218, USA}
\affil{$^3$Deborah Lunder and Alan Ezekowitz Founders' Circle Member, Institute for Advanced Study, Einstein Dr., Princeton, NJ 08540, USA} 
\affil{$^4$Canadian Institute for Theoretical Astrophysics, 60 St. George St, University of Toronto, Toronto, ON M5S 3H8, Canada; \\
Dunlap Institute for Astronomy and Astrophysics, The University of Toronto, Toronto, ON M5S 3H4, Canada}
\affil{$^5$N\'ucleo de Astronom\'{\i}a de la Facultad de Ingenier\'{\i}a, Universidad Diego Portales, Av. Ej\'ercito Libertador 441, Santiago, Chile}
\affil{$^6$Institute of Astronomy, University of Cambridge, Madingley Road, Cambridge CB3 0HA, UK}  
\affil{$^7$Department of Physics \& Astronomy, University of California, Riverside, CA 92507, USA}
\affil{$^8$Department of Astronomy \& Astrophysics, The Pennsylvania State University, University Park, PA 16802, USA}
\affil{$^9$Centre for Extragalactic Astronomy, Department of Physics, Durham University, South Road, Durham, DH1 3LE, UK} 
\affil{$^{10}$European Southern Observatory, Karl Schwarzschild Stra{\ss}e 2, 85748 Garching bei M{\"u}nchen, Germany}
\affil{$^{11}$Aix-Marseille Universit\'e, CNRS, Laboratoire d'Astrophysique de Marseille, UMR 7326, 13388 Marseille, France}
\affil{$^{12}$Department of Physics, Drexel University, 3141 Chestnut Street, Philadelphia, PA 19104, USA}
\affil{$^{13}$Jet Propulsion Laboratory, California Institute of Technology, 4800 Oak Grove Drive, Mail Stop 169-236, Pasadena, CA 91109, USA}

\begin{abstract}
Quasars may have played a key role in limiting the stellar mass of massive galaxies. Identifying those quasars in the process of removing star formation fuel from their hosts is an exciting ongoing challenge in extragalactic astronomy. In this paper we present X-ray observations of eleven extremely red quasars (ERQs) with $L_{\rm bol}\sim 10^{47}$ erg~s$^{-1}$ at $z=1.5-3.2$ with evidence for high-velocity ($v\ga 1000$ km~s$^{-1}$) [OIII]$\lambda$5007\AA\ outflows. X-rays allow us to directly probe circumnuclear obscuration and to measure the instantaneous accretion luminosity. We detect ten out of eleven extremely red quasars available in targeted and archival data. Using a combination of X-ray spectral fitting and hardness ratios, we find that all of the ERQs show signs of absorption in the X-rays with inferred column densities of $N_{\rm H}\approx 10^{23}$ cm$^{-2}$, including four Compton-thick candidates ($N_{\rm H}\ga 10^{24}$ cm$^{-2}$). We stack the X-ray emission of the seven weakly detected sources, measuring an average column density of $N_{\rm H}\sim 8\times 10^{23}$ cm$^{-2}$. The absorption-corrected (intrinsic) $2-10$ keV X-ray luminosity of the stack is $2.7\times 10^{45}$ erg~s$^{-1}$, consistent with X-ray luminosities of type 1 quasars of the same infrared luminosity. Thus, we find that ERQs are a highly obscured, borderline Compton-thick population, and based on optical and infrared data we suggest that these objects are partially hidden by their own equatorial outflows. However, unlike some quasars with known outflows, ERQs do not appear to be intrinsically underluminous in X-rays for their bolometric luminosity. Our observations indicate that low X-rays are not necessary to enable some types of radiatively driven winds. 
\end{abstract}


\section{Introduction}
\label{sec:intro}

It has been known for decades that the majority of present-day galaxies host a supermassive black hole and that accretion onto these black holes is the source of quasar emission. Based on the demographics, polarimetry and X-ray observations of active nuclei, the unification model \citep{anto93} posits that obscured (type 2) quasars are seen through a torus of gas and dust but would otherwise appear as classic unobscured type 1 quasars. At the same time, theoretical models of galaxy formation suggest that obscured quasars may represent an early phase of active black hole growth in which galaxy-wide obscuration leads to the observed properties \citep{sand96, cana01, hopk06}. In such models, rapidly accreting obscured black holes can drive winds that clear the galaxy of gas, shutting off star formation in the process now commonly referred to as quasar feedback \citep{silk98, king03}. Thus, obscuration can be both due to geometric orientation and/or through evolutionary effects, and the role of these factors in quasar demographics and galaxy evolution is potentially important and remains poorly understood. 

Until recently, direct observations of galactic-scale quasar-driven winds have been scarce, but in the last few years this multi-phase phenomenon has been observed via ultraviolet and X-ray absorption (e.g., \citealt{hama01, moe09, nard15}), optical line emission (e.g., \citealt{gree11, cano12, liu13a, liu13b, harr14}) and molecular transitions (e.g., \citealt{veil13a, fior17}). One conclusion emerging from this work is that in quasars with powerful galaxy-wide outflows, the classical ``narrow-line" region of quasars is no longer confined by the potential of the host galaxy and exhibits high velocity dispersions, with full width at half maximum (FWHM) of the forbidden emission lines (e.g., [OIII]$\lambda$5007\AA) of $\ga 1000$ km s$^{-1}$, and blue-shifted asymmetries \citep{liu13b}. Such criteria on line shapes and widths are widely used for identifying quasars and galaxies with outflows (e.g., Mullaney et al. 2013; \citealt{brus15}). 

The period around $z\sim 2-3$ is particularly important in galaxy formation because it marks the peak of both star formation and quasar activity in the universe \citep{boyl98}. However, identifying quasars exhibiting feedback at high redshifts is challenging. In unobscured quasars, the emission signatures of winds are difficult to detect in proximity to the bright central source. Furthermore, feedback may be primarily associated with obscured (and thus optically faint) sources which are yet to be identified in large numbers. Over the last several years, our group has developed a range of approaches to identify obscured and reddened luminous quasars at $1.5<z<4$ \citep{alex13, gree14b, ross15, hama17} using data from the Sloan Digital Sky Survey (SDSS; \citealt{eise11, daws13}) and the Wide-Field Infrared Survey Explorer (WISE; \citealt{wrig10}). In particular, extremely red quasars (ERQs), selected on the basis of high infrared-to-optical ratios and high equivalent width of CIV$\lambda$1549\AA, often show signs of extreme outflow activity in their [OIII]$\lambda\lambda$4959,5007\AA\ emission, unmatched by any other quasar sample \citep{zaka16b, perr17}. Because [OIII] likely traces relatively low density gas, these outflows may be occurring on galaxy-wide scales, and therefore these objects could be manifestations of quasars during a strong feedback episode. In order to better constrain the intrinsic power of these quasars, and assess whether they are capable of driving large-scale outflows, we require a direct measurement of their accretion luminosities. Even in the presence of large absorbing column densities, such as those expected in ERQs, hard X-ray emission arising from the quasar provides a robust determination of the bolometric luminosity.

In this paper we present the first results of our follow-up X-ray programs of this intriguing population of ERQs. In Section \ref{sec:data}, we describe the sample and the observations. In Section \ref{sec:analysis}, we present X-ray spectroscopic analysis. We discuss our results in Section \ref{sec:disc} and conclude in Section \ref{sec:conc}. We adopt an $h=0.7$, $\Omega_{\rm m}=0.3$, $\Omega_{\Lambda}=0.7$ cosmology and identify optical emission lines using their wavelengths in air following long-standing usage. Objects are identified as SDSS~Jhhmmss.ss+ddmmss.s in Table \ref{tab:all} and as SDSS~Jhhmm+ddmm elsewhere. {\bf W1, W2, W3 and W4} refer to the $3.6\micron$, $4.5\micron$, $12\micron$, and $22\micron$ Vega-based magnitudes from the AllWISE data release \citep{wrig10,cutr13}, and $u,g,r,i,z$ to AB-based magnitudes from the SDSS \citep{eise11,alam15}.

\section{Observations and data reduction}
\label{sec:data}

\subsection{Parent ERQ samples and target selection}

\begin{figure}
\includegraphics[width=\linewidth]{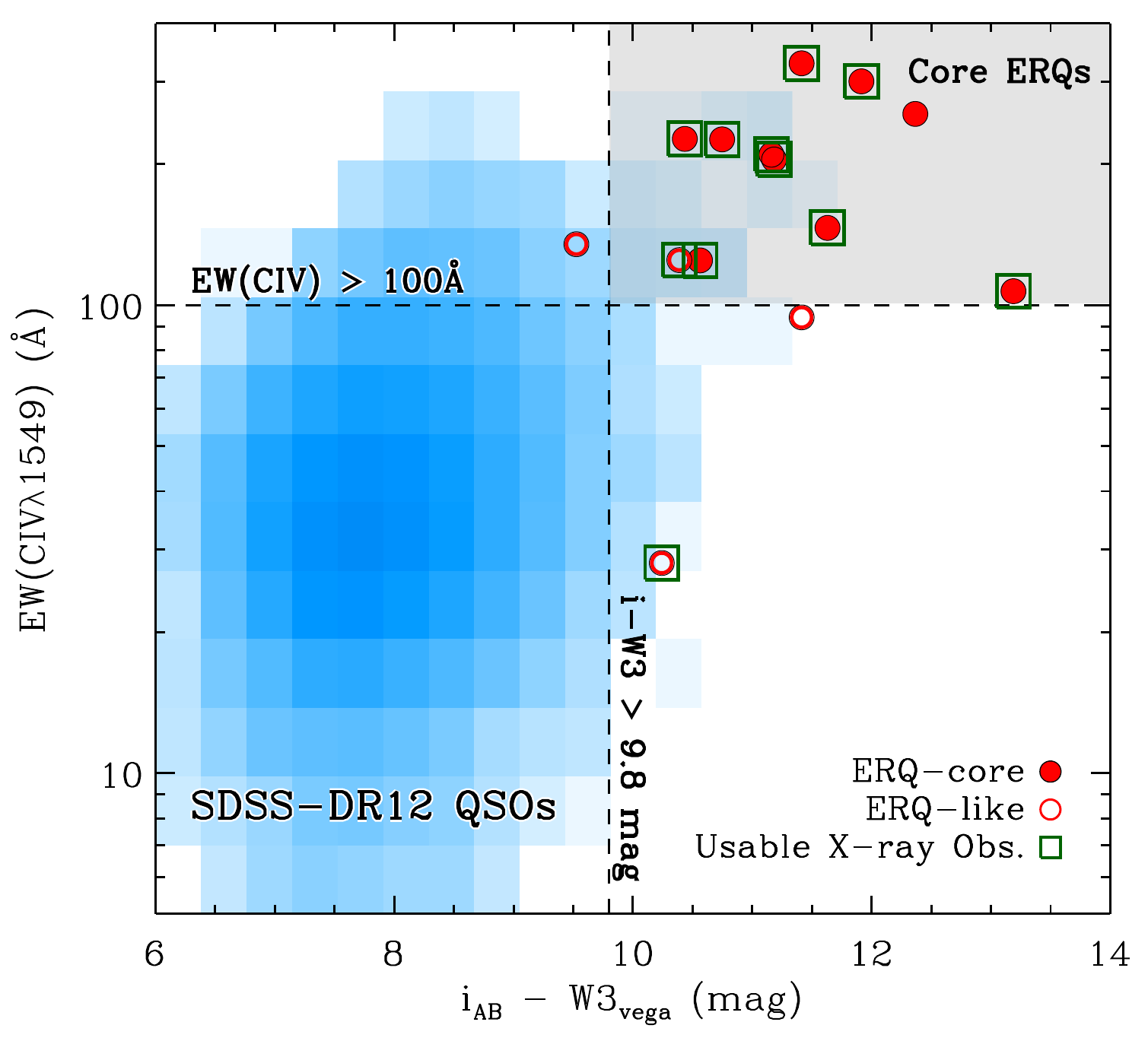}
\caption{Target selection in the space of optical-to-infrared color and the CIV equivalent width. Shaded contours show the locus of all SDSS-DR12 quasars between $2<z<3$ that are detected in the WISE W3 band (S/N~$> 2$). ERQ-core and ERQ-like objects presented in this paper are shown with filled and open circles, respectively. Those objects covered by X-ray observations that we determined to be of sufficient quality for our purposes are highlighted with green squares. The gray shaded box highlights the region used by \citet{hama17} to describe the ERQ-core sources.}
\label{pic:selection}
\end{figure}

Extremely red quasars (ERQs) were first identified based on their extremely high IR-to-optical flux ratios, with $r_{\rm AB}-{\rm W4}_{\rm Vega} > 14$ mag \citep{ross15} within a sample of spectroscopically confirmed, optically-selected quasars \citep{pari14, pari17} in the Baryon Oscillation Spectroscopic Survey (BOSS; \citealt{daws13}). In addition to their extreme infrared-to-optical colors, some of these objects showed unusual emission line properties. In particular, they display a mix of ``type 1'' (unobscured) characteristics -- such as broad emission lines (FWHM of CIV$\lambda$1550\AA\ is $>$2000 km~s$^{-1}$) -- and ``type 2'' (obscured) characteristics, such as high equivalent widths of emission lines, indicative of continuum suppression. 

\citet{hama17} re-examined and formalized the ERQ selection to include both photometric and spectroscopic selection criteria. Using a color cut of $i_{\rm AB}-{\rm W3}_{\rm Vega}>9.8$ mag {\it and} rest equivalent width of CIV$\lambda$1550\AA\ $>$100\AA, \citet{hama17} identified a sample of 97 ERQs at $2<z<3$, hereafter referred to as our parent `ERQ-core' sample (shaded region of Fig.1). Based on further analysis of the SDSS DR12 quasar sample, \citet{hama17} defined a second sample of 235 quasars that exhibited any of the following characteristics in the broadband photometry or spectroscopy: $i_{\rm AB}-{\rm W3}_{\rm Vega}>9.8$ mag, or ${\rm EW(CIV)} > 100$\AA, or CIV line profiles that are ``wingless" or ``boxy" (quantified by the line profile's kurtosis) with $kt_{80} > 0.33$. Specifically, $kt_{80}$ measures the velocity width of the emission line profile at 80\% of its peak height divided by the width at 20\%. These objects, which display some, but not all, of the properties of the core ERQs, are hereafter referred to as our parent `ERQ-like' sample (Section 5.7 and Table B1 of \citealt{hama17}). The distribution and selection of the ERQ samples as compared to the SDSS-DR12 quasars are shown in Figure \ref{pic:selection}. One source in our sample, SDSS J212951.40$-$001804.3, was determined by \citet{hama17} to be ERQ-like as it was not detected in the version of the WISE source catalog used in their previous analysis. In the most recent WISE release this source is detected in W3 with S/N$\sim 4.7$, and here we use its updated WISE photometry, $W3_{\rm Vega} = 11.407$ mag, which would now reclassify the source as an ERQ-core object. However, for ease of comparison to the \citet{hama17} catalog, we show SDSS J212951.40$-$001804.3 as ERQ-like in Figure \ref{pic:selection}.

We are conducting an extensive campaign to obtain multi-wavelength follow-up of the ERQ samples, including radio observations \citep{hwan17}, optical spectropolarimetry \citep{alex17}, and extensive near-infrared spectroscopy with the VLT, Gemini and Keck \citep{zaka16b, perr17}. The near-infrared spectra probe rest-frame optical wavelengths and cover the key diagnostics H$\beta$+[OIII]$\lambda\lambda$4959,5007\AA\AA\ and H$\alpha$+[NII]$\lambda\lambda$6548,6563\AA\AA. Our follow-up near-infrared spectroscopy of $\sim 20$ sources reveals that ERQs routinely show kinematically disturbed [OIII] emission lines \citep{perr17}, inconsistent with gas in a galactic potential and passively photo-ionized by the quasar, FWHM reaching an unprecedented $>$5000 km s$^{-1}$ in some sources \citep{zaka16b, perr17}. 

\begin{figure*}
\includegraphics[width=\linewidth]{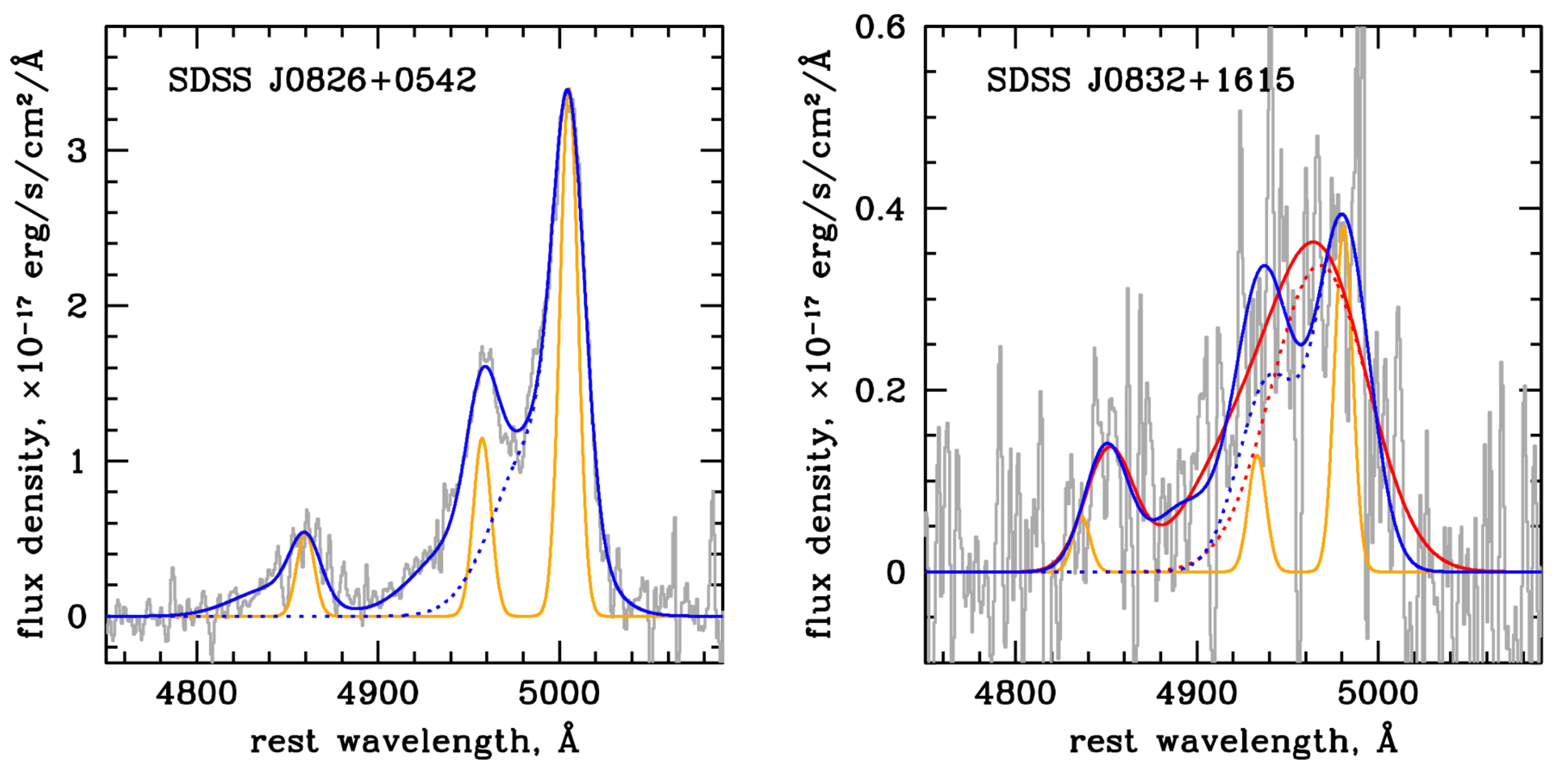}
\caption{Gemini GNIRS [OIII]$\lambda\lambda$4959,5007\AA\AA+H$\beta$ spectroscopy (PI Alexandroff) of two of the objects presented here. Solid lines show the overall fit (blue for two-Gaussian fits, red for one-Gaussian fits), and dotted lines show the [OIII]$\lambda$5007\AA\ contribution separately. In SDSS~J0826+0542, outflow signatures include a blue-shifted broad wing of [OIII] and the large velocity width of the entire line (FWHM[OIII]$\simeq 1600$ km s$^{-1}$). SDSS~J0832+1615 is much fainter, so the spectrum has a lower signal signal-to-noise ratio, but [OIII]$\lambda$4959\AA\ and [OIII]$\lambda$5007\AA\ are strongly blended, suggesting velocity FWHM$\simeq 3800$ km s$^{-1}$ in both the single-Gaussian (red) and the double-Gaussian (blue) fitting attempts. For comparison, orange line shows [OIII]$+$H$\beta$ profiles with kinematically undisturbed FWHM$=700$ km s$^{-1}$.}
\label{pic:spec}
\end{figure*}

As an example, in Figure~\ref{pic:spec}, we show previously unpublished Gemini GNIRS spectra of two objects from our X-ray sample presented in this paper, where we illustrate our identification of [OIII] wind signatures. Details of observations and fitting are provided in \citet{alex17}. We find large velocity widths of [OIII] in both objects, with FWHM in the top 1\% of the [OIII] widths of the low-redshift obscured quasar population \citep{zaka14,yuan16}. Three of the remaining four near-infrared spectra mentioned in Table \ref{tab:all} are published in \citet{zaka16b} and \citet{alex17}. A complete analysis of our multi-facility near-infrared spectroscopic campaign will be presented by \citet{perr17}. While it is not yet clear exactly which properties of ERQs -- their luminosities, their colors, or their rest-frame UV line shapes -- are most strongly associated with the kinematic activity in [OIII], ERQs selected on the basis of colors and CIV equivalent widths show near-100\% detection rate of [OIII] outflows \citep{perr17}.

\subsection{ERQs with sensitive X-ray observations}

In 2014, we started follow-up X-ray observations of $1.5<z<3$ ERQs from \citet{ross15} and \citet{hama17} with confirmed [OIII] outflows. We proposed for three objects to be observed with XMM-Newton (PI Alexandroff) in Cycle 14. Two were observed (SDSS J0834+0159; SDSS J2323-0100) by XMM. Seven further ERQ targets were proposed for with Chandra (PI Zakamska) in Cycle 17. Four were approved and observed. The observed targets are listed in Table \ref{tab:all}, and the distribution of targets in the space of infrared-to-optical color and CIV equivalent width -- the defining properties of ERQs -- is shown in Figure \ref{pic:selection}. Furthermore, the results of our near-infrared spectroscopy for the six targeted sources are also listed in Table \ref{tab:all}. By selection, all six of our observed X-ray targets show [OIII] with strong wind signatures, with FWHM of [OIII] between 1500$-$2800 km~$^{-1}$.

In addition to our own targeted observations, we include archival X-ray observations from XMM and Chandra in our parent ERQ-core and ERQ-like samples. We use NASA's High Energy Astrophysics Science Archive Research Center (HEASARC) tools to cross-correlate all 97 ERQ-core and 235 ERQ-like quasars against the \texttt{xmmmaster} catalog (which contains the coordinates of all XMM pointings) within 2\arcmin. Outside of this radius, the point-spread function (PSF) becomes highly non-circular making weak sources difficult to accurately identify. Further, we cross-match the sample against the XMM source catalog (\texttt{xmmssc}, \citealt{rose16}) for ERQs at larger off-axis distances. We identified two ERQs (SDSS~J0006+1215; SDSS~J1700+4002) with on-axis XMM observations, and a further two ERQs (SDSS~J0220+0137; SDSS~J0116$-$0505) identified within the \texttt{xmmssc}. Of the two on-axis XMM sources, one was targeted as a comparison object to another sample of reddened quasars \citep{bane15}, and the other is serendipitously close to the center of an unrelated observation. 

To identify any ERQs that overlap with existing Chandra observations, instead of the Chandra Source Catalog (\texttt{cxogsgsrc}, \citealt{wang16a}) whose detection threshold is too high for faint sources such as ours \citep{goul12}, we use the {\tt find\_chandra\_obsid} tool. Specifically, we include only those observations that cover ERQ positions on Chips I0--3 in ACIS-I mode or Chip S3 in ACIS-S mode, where adverse PSF effects of Chandra are reduced. We identified a further four ERQs that met these criteria -- two were targeted as hot dust-obscured galaxy candidates (HotDOGs; Section \ref{sec:hotdogs}), and the remaining two are off-axis serendipitous observations.

Our final sample contains 14 ERQ-core or ERQ-like objects with X-ray observations (Table \ref{tab:all}), though only 11 ERQs have usable X-ray data (Table \ref{tab:xray}) as discussed in the next Section. Our selection does not include quasars which might be covered by off-axis XMM observations if they are not strongly detected in X-rays (i.e., they must be included in the \texttt{xmmssc}). Thus, our matching strategy is marginally incomplete and partially biased toward brighter X-ray sources, in that we are not considering the upper limits on X-ray fluxes that might be available for such objects.

\begin{table*}
\begin{center}
\setlength{\tabcolsep}{0.9mm}
\caption{High-redshift extremely red quasars with X-ray observations, by optical selection and observation type \label{tab:all}}
\footnotesize
\begin{tabular}{lrrccccccccccccc}
\hline\hline
  \multicolumn{1}{c}{SDSS ID} &
  \multicolumn{1}{c}{$\alpha$} &
  \multicolumn{1}{c}{$\delta$} &
  \multicolumn{1}{c}{$z$} &
  \multicolumn{1}{c}{Class} &  
  \multicolumn{1}{c}{[O{\sc iii}]} &
  \multicolumn{1}{c}{Outflow?} &
  \multicolumn{1}{c}{X-ray} &
  \multicolumn{1}{c}{Inst.} &
  \multicolumn{1}{c}{$t_{\rm exp}$} &
  \multicolumn{1}{c}{$d_{\rm OAX}$} &
  \multicolumn{1}{c}{$\nu L_{6 \mu m}$} \\
  \multicolumn{1}{c}{} &
  \multicolumn{1}{c}{(deg)} &
  \multicolumn{1}{c}{(deg)} &
  \multicolumn{1}{c}{} &
  \multicolumn{1}{c}{} &  
  \multicolumn{1}{c}{spec?} &
  \multicolumn{1}{c}{} &
  \multicolumn{1}{c}{Telescope$^\dagger$} &
  \multicolumn{1}{c}{} &
  \multicolumn{1}{c}{(ks)} &
  \multicolumn{1}{c}{($'$)} &
  \multicolumn{1}{c}{(log erg/s)} \\
  \multicolumn{1}{c}{(1)} &
  \multicolumn{1}{c}{(2)} &
  \multicolumn{1}{c}{(3)} &
  \multicolumn{1}{c}{(4)} &
  \multicolumn{1}{c}{(5)} &  
  \multicolumn{1}{c}{(6)} &
  \multicolumn{1}{c}{(7)} &
  \multicolumn{1}{c}{(8)} &
  \multicolumn{1}{c}{(9)} &
  \multicolumn{1}{c}{(10)} &
  \multicolumn{1}{c}{(11)} &
  \multicolumn{1}{c}{(12)} \\
\hline
SDSS J000610.67+121501.2   &   1.54446 & 12.25033 & 2.309 & ERQ      & \xmark & ?      & XMM     & PN,M1/2 & 23 & on-axis & 47.04 \\
SDSS J011601.43$-$050503.9 &  19.00596 & -5.08442 & 3.183 & ERQ-like & \xmark & ?      & Chandra & ACIS-S  & 71 & on-axis & 47.25 \\
SDSS J022052.13+013711.4   &  35.21721 &  1.61983 & 3.138 & ERQ      & \xmark & ?      & Chandra & ACIS-S  & 70 & on-axis & 47.26 \\
SDSS J082653.42+054247.3   & 126.72258 & 5.71314  & 2.578 & ERQ      & \cmark & \cmark & Chandra & ACIS-S  & 15 & on-axis & 46.78 \\ 
SDSS J083200.20+161500.3   & 128.00083 & 16.25008 & 2.431 & ERQ      & \cmark & \cmark & Chandra & ACIS-S  & 15 & on-axis & 46.72 \\ 
SDSS J083448.48+015921.1   & 128.70200 & 1.98919  & 2.591 & ERQ      & \cmark & \cmark & XMM     & PN,M1/2 & 35 & on-axis & 47.03 \\
SDSS J091508.45+561316.0   & 138.78521 & 56.22111 & 2.857 & ERQ      & \xmark & ?      & Chandra & ACIS-S  & 23 & 5.0     & 46.69 \\
SDSS J112124.55+570529.6   & 170.35229 & 57.09156 & 2.383 & ERQ-like & \xmark & ?      & Chandra & ACIS-I  & 5  & 4.6     & 46.80 \\
SDSS J131047.78+322518.3   & 197.69908 & 32.42175 & 3.009 & ERQ      & \xmark & ?      & XMM     & M1/2    & 50 & 11.4    & 47.12 \\
SDSS J153542.40+090341.1   & 233.92667 & 9.06142  & 1.533 & ERQ      & \cmark & \cmark & Chandra & ACIS-S  & 15 & on-axis & 46.40 \\
SDSS J165202.64+172852.4   & 253.01100 & 17.48122 & 2.942 & ERQ      & \cmark & \cmark & Chandra & ACIS-S  & 15 & on-axis & 47.19 \\
SDSS J170047.07+400238.7   & 255.19613 & 40.04408 & 2.903 & ERQ-like & \xmark & ?      & XMM     & PN,M1/2 & 11 & 1.8     & 46.39 \\
SDSS J212951.40$-$001804.3 & 322.46417 & -0.30119 & 3.206 & ERQ-like & \xmark & ?      & XMM     & PN,M1/2 & 34 & 9.3     & $<$46.56 \\
SDSS J232326.17$-$010033.1 & 350.85904 & -1.00919 & 2.356 & ERQ      & \cmark  & \cmark & XMM     & PN,M1/2 & -  & on-axis & 46.58 \\
\hline
\end{tabular}
\end{center}
$^1$Full SDSS-DR13 designation;
$^{2-3}$J2000 coordinates in degrees;
$^4$Redshift; 
$^5$Source class based on the ERQ and ERQ-like definitions outlined in Section~\ref{sec:data}; 
$^6$Does the source have available rest-frame optical spectroscopy covering the [OIII] emission line at 5007\AA?; 
$^7$If the source has available [OIII] spectroscopy, is the [OIII] observed to be strongly asymmetric and determined to host an outflow?; 
$^{8-9}$X-ray telescope and associated instruments for the source; 
$^{10}$Exposure time in kiloseconds for the X-ray observation; 
$^{11}$Off-axis distance in arcminutes for the position of the source on the X-ray detector; 
$^{12}$Rest-frame $6_{\mu m}$ luminosity derived from the WISE infrared photometry.
$^\dagger$All sources observed with Chandra are included in the X-ray stacking analysis presented in Section~3.3.

\end{table*}

\subsection{X-ray data reductions}

The data for the six ERQ targets observed by XMM-Newton were retrieved from the HEASARC database and reduced using standard tools in the XMM-Newton {\sc sas} software, version 15.0.0. Specifically, we apply the latest calibration files and analyze the light curves for flaring events for the MOS and PN CCDs from the European Photon Imaging Camera (EPIC). We determine that each of the targets covered by XMM observations was subject to some X-ray flaring, typically at the $\sim$10--20\% level of the total observation exposure time. These periods of flaring are identified and removed from the observations. For SDSS~J1700+4002 the entire 11~ks X-ray exposure was subject to high levels of flaring, resulting in no scientifically usable data. We remove this object from our sample and do not discuss it further. Final flare-corrected exposure times are given in Table \ref{tab:xray} and are in the range 12--50~ks.

Photon events are extracted from elliptical apertures based upon the size and shape of the PN/MOS PSF at the detector position of the target. As two of the targets were serendipitously detected within the XMM observations, they are positioned at large off-axis angles, which significantly distorts the shape of the PSF from that of a circle, which we account for based on XMM PSF models. For the two sources with $> 100$ photon counts in the 0.5-10~keV band, we group the counts into bins of 15 counts for $\chi^2$ statistics and construct aspect histograms and response matrices at the positions of the targets in order to derive the final X-ray spectra. Finally, the X-ray emission from SDSS~J2323-0100 is heavily contaminated by an unexpectedly bright foreground quasar (angular separation $\sim$15\arcsec). Because of the low spatial resolution of the XMM observation, this observation is unusable and we remove this object from further analyses. 

Of the eight ERQs observed with {\it Chandra}, six were observed on-axis with ACIS-S, but one of them (SDSS~J0116$-$0505) is not yet public and is therefore excluded from all analyses. An additional source was serendipitously identified in an ACIS-S observation, $\sim$5 arc-minutes off-axis on the ACIS-S3 chip, and the last Chandra source was observed $\sim$4.6 arc-minutes off-axis with ACIS-I. The individual observation identification numbers (ObsIDS) and the X-ray fluxes and counts are provided in Table \ref{tab:xray}. For each of the Chandra observations, we carry out data processing using the Chandra X-ray Center software packages available in {\sc ciao} v4.8 in conjunction with the latest calibration files ({\sc caldb} 4.7.3) applied using {\tt chandra\_repro}. Streak events, bad pixels, pixel randomization and cosmic rays are removed with STATUS=0 and screened with the typical grad set during the implementation of {\tt acis\_process\_events}. Flares greater than 3$\sigma$ above the background are identified and removed using {\tt lc\_clean} to create Level-2 events files. Flare corrected exposure times for the {\it Chandra} observations are in the range $\sim 5$--30~ks. Point source photometry is performed using circular apertures based on the size of a region required to enclose 90\% of the PSF ($r_{\rm 90}$) at 0.5, 2 and 5~keV, with background regions defined by annuli with outer radii equal to $5r_{\rm 90}$ and inner radii equal to 1.3$r_{\rm 90}$. Counts are extracted and exposure maps are constructed through the implementation of the {\tt srcflux} tool in {\sc ciao} for energy bands 0.3--1, 1--4 and 4--7~keV (see Section~\ref{sec:det}).

\section{X-ray spectral analysis}
\label{sec:analysis}

We identify X-ray detections and measure X-ray fluxes and spectral shapes in Section \ref{sec:det} and summarize these results in Table \ref{tab:xray}. In Section \ref{sec:ind} we discuss some of the sources individually. In Section \ref{sec:stack} we stack the Chandra sources -- which are too weakly detected for individual spectral fits -- and conduct a spectral analysis of the stack. 

\subsection{X-ray flux and hardness measurements}
\label{sec:det}

\begin{figure*}
\includegraphics[scale=0.87,trim=0cm 0cm 0cm 0cm,clip=true]{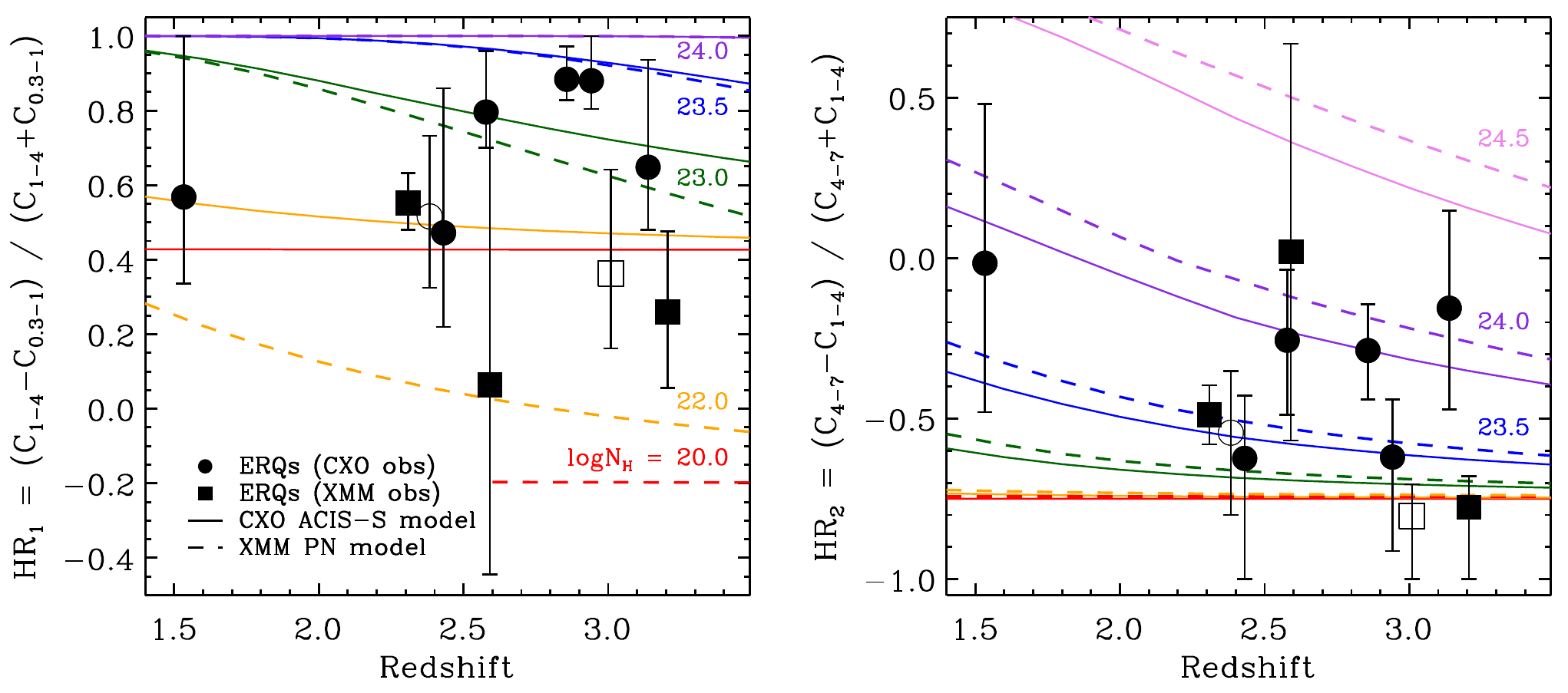}
\caption{Observed photon hardness ratios versus redshift for our sample of ERQ sources with ${\rm HR}_1 = (C_{\rm 1-4 keV} - C_{\rm 0.3-1keV})/(C_{\rm 1-4 keV} + C_{\rm 0.3-1keV})$ (left figure) and ${\rm HR}_2 = (C_{\rm 4-7 keV} - C_{\rm 1-4keV})/(C_{\rm 4-7 keV} + C_{\rm 1-4keV})$ (right figure). HRs are computed using the Bayesian Estimation of Hardness Ratios (BEHR) software package. Uncertainties are derived from 67\% of the posterior samples. Filled (open) circles represent sources observed with Chandra ACIS-S (ACIS-I) and filled (open) squares are those ERQs observed with XMM-PN (MOS1/2). Model HRs obtained for an absorbed power-law spectrum with slope $\Gamma=1.8$ blocked by a range of redshifted column densities ($N_{\rm H}$) are shown for XMM-PN (dashed lines) and Chandra ACIS-S (solid lines) detectors.}
\label{pic:hr}
\end{figure*}

We divide the observed energy range into three bands: soft (0.3$-$1 keV), middle (1$-$4 keV) and hard (4$-$7 keV). At $z\sim 2.5$, this provides us with rest-frame energy ranges of $\sim 1-3.5$, $\sim 3.5-14$, and $\sim 14-24$ keV. All seven sources observed with Chandra were detected in at least one energy band, and similarly, 3 of 4 sources were detected with XMM. Only 4 of the 11 objects were detected in the soft band, two of each in {\it Chandra} and {\it XMM}. We detect all seven of the {\it Chandra} objects in the middle band, with four of these also detected in the hard band. We detected 3(2) out of 4 sources in the middle(hard) bands with XMM. The majority of the detected sources are observed with $S/N > 5$ above the background in the middle band with either {\it Chandra} or {\it XMM}. Given the relatively short {\it Chandra} exposures, the expected background counts in the source apertures in the middle band are typically $<1$~count. The object most weakly detected with {\it Chandra} (SDSS J1535+0903) has 3 counts at $E \sim 1-7$~keV. More typically, our sources have 5--15 counts. The individual source counts and fluxes (or 3$\sigma$ upper limits) are listed in Table~\ref{tab:xray}. For consistency between the telescopes, we have chosen to limit both the Chandra and XMM data to 7~keV for all of our energy-band analyses. Even though the effective area of XMM extends beyond 7~keV, we find that limiting the energy range in this manner does not adversely affect our conclusions given that the two sources observed by XMM that do not have the required counts to construct a spectrum are already not detected in the 4--7 keV range, and hence, we are not excluding potential photons at 7--10~keV from our analyses.

It is conventional to quote X-ray luminosities over the rest-frame $2-10$ keV range ($L_{\rm 2-10 keV}$ hereafter). For a power-law X-ray spectrum with a typical slope $\Gamma=1.8$ (${\rm d}N/{\rm d}E\propto E^{-\Gamma}$, \citealt{nand94,reev00,pico05,page05}), and in the absence of obscuration, the ratio of the $3.5-14$ keV to the $2-10$ keV luminosities is close to unity, making our middle band well-suited for the determination of X-ray luminosity.  As the cross-section for photo-electric absorption steeply declines with energy, soft X-ray photons are absorbed by the intervening material a lot more effectively than the hard X-ray photons. As a result, for a Compton-thick absorber with $N_{\rm H} \sim 3 \times 10^{24}$~cm$^{-2}$ the ratio of the apparent $3.5-14$ keV to the $2-10$ keV luminosities is $\sim 10$. We are interested in the intrinsic absorption-corrected X-ray luminosities, so the redshift of the sample works in our favor in moving the relatively less absorbed parts of the spectrum into the observable energy range. Nonetheless, to determine the intrinsic luminosities we need measurements of absorption and calculations of absorption corrections.   

To gauge whether absorption is important in our sample we first look at the flux ratios. For a power-law spectrum with index $\Gamma\ne 2$, the flux ratio between two bands from $E_1-E_2$ and from $E_2-E_3$ is $F_{E_2-E_3}/F_{E_1-E_2}=(E_3^{2-\Gamma}-E_2^{2-\Gamma})/(E_2^{2-\Gamma}-E_1^{2-\Gamma})$. For standard values of $\Gamma=1.65-1.95$ the expected hard-to-middle band flux ratio is $F_{\rm 4-7 keV}/F_{\rm 1-4 keV}=0.42-0.56$, where a higher ratio would indicate lower values of $\Gamma$ and harder spectra. The six detected objects have much harder spectra, with a median $F_{\rm 4-7 keV}/F_{\rm 1-4 keV}\simeq 1.2$, and the four remaining objects with hard-band upper limits consistent with this value. Only one source -- SDSS~J0006+1215 -- with the hard-to-middle flux ratio of 0.36 is marginally consistent with a power-law spectrum with $\Gamma\ga 1.95$. 

The most natural explanation for the hardness of the observed spectra is photo-electric absorption, strong enough to suppress rest-frame $3.5-14$ keV flux. Measurements of the photo-electric absorption due to intervening gas typically require detailed modeling of the observed X-ray spectrum. However, the majority of the ERQs are only weakly detected in the X-rays and lack the necessary counts to perform such an analysis. In the absence of detailed X-ray spectroscopy, we can estimate the obscuring column density ($N_{\rm H}$) through the use of X-ray hardness ratios (HR). HRs compare the observed photon counts between two bands, with the harder (higher energy) photons being less affected by photoelectric absorption than the softer photons. Here we use two sets of hardness ratios derived from the 0.3--1~keV and 1--4~keV bands and the 1--4~keV and 4--7~keV bands, defined as:

\begin{equation}
{\rm HR}_1=\frac{C_{\rm 1-4 keV}-C_{\rm 0.3-1 keV}}{C_{\rm 1-4 keV}+C_{\rm 0.3-1 keV}},
\end{equation}

\begin{equation}
{\rm HR}_2=\frac{C_{\rm 4-7 keV}-C_{\rm 1-4 keV}}{C_{\rm 4-7 keV}+C_{\rm 1-4 keV}}.
\end{equation}
Defined in this way for $z=2.5$ targets, these values are roughly equivalent to the standard HR bands used for analysis of local systems with Chandra/XMM in the case of HR$_1$ or NuSTAR/Suzaku for HR$_2$.

\begin{figure*}
\includegraphics[scale=0.58,trim=2cm 13cm 4cm 1cm,clip=true]{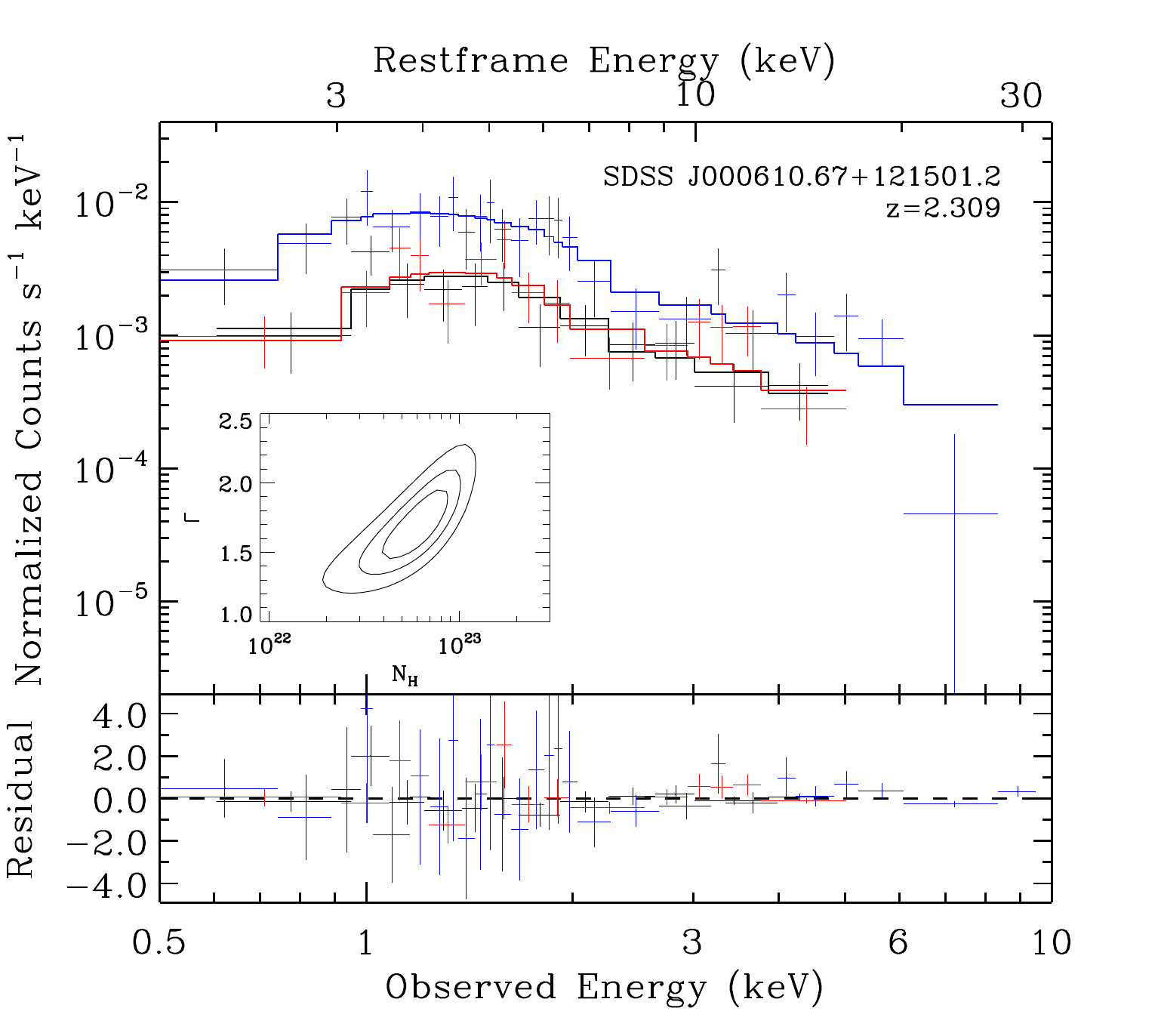}%
\includegraphics[scale=0.58,trim=2cm 13cm 4cm 1cm,clip=true]{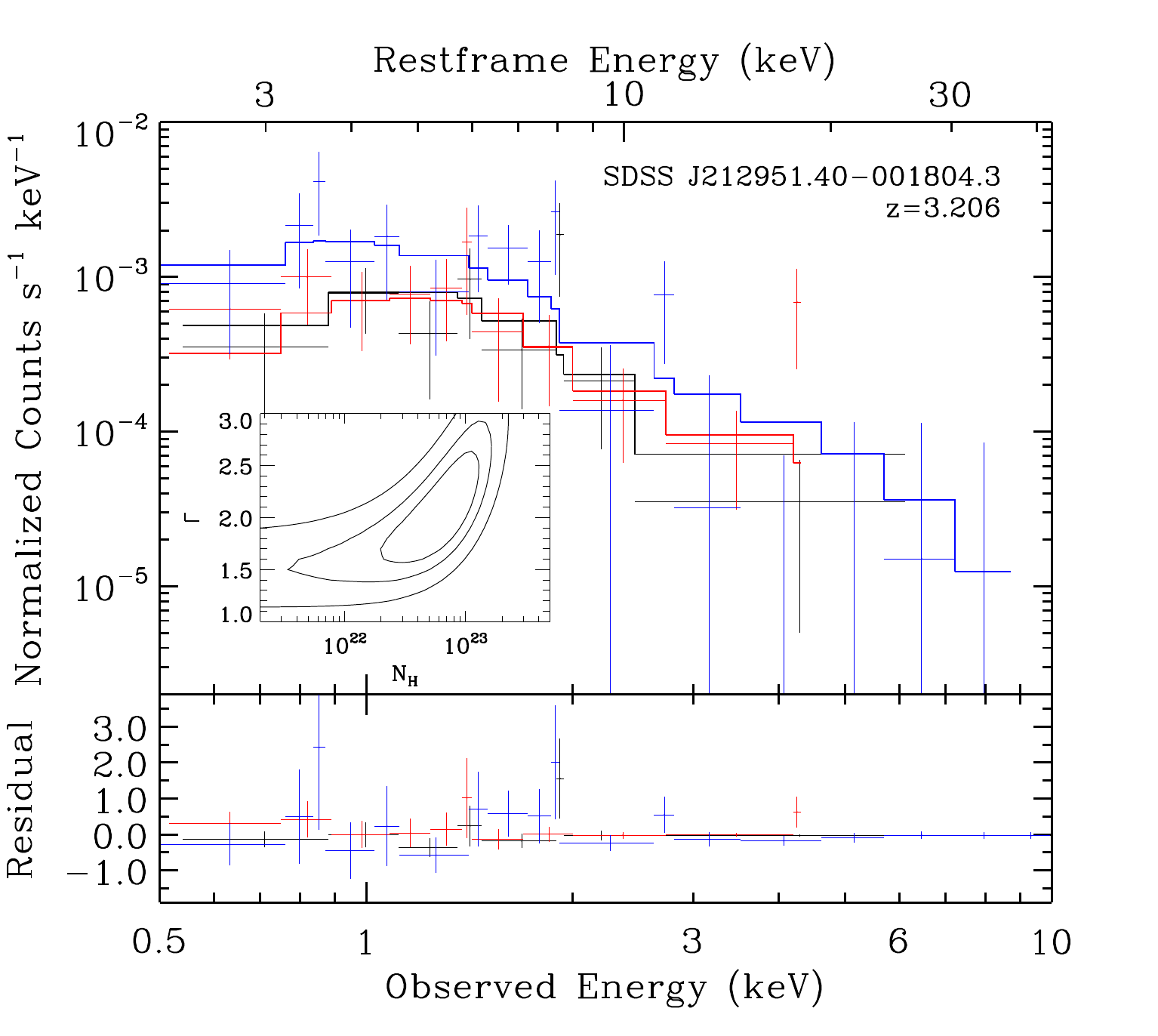}
\caption{Top panels: XMM-Newton spectra for SDSS J0006+1215 (left) and SDSS J2129-0018 (right). Spectra are binned to have at least 15 counts in each bin. EPIC-PN and MOS1/2 are shown with blue, black and red symbols, respectively. Insets provide the 1,2 and 3$\sigma$ contours for the $N_{\rm H}$ and spectral slope ($\Gamma$) parameters produced from a simple rest-frame absorbed power-law model that has been fit to the data. Lower panels provide the residual between the best-fit model and the data.}
\label{pic:xray}
\end{figure*}

We use the Bayesian Estimation of Hardness Ratios (BEHR) software package \citep{park06} to measure the HRs and estimate their uncertainties for all of the ERQs in our sample. BEHR computes posterior draws for the HRs assuming each detected photon (source or background) is an independent Poisson random variable. It is particularly useful in the low count Poisson regime as it is capable of producing posterior draws irrespective of whether the source is formally detected in both energy bands. As is recommended with BEHR when computing HRs, we assume uniform priors and employ a Gaussian quadrature algorithm to compute the Markov-Chain Monte Carlo integrals with the default 1000 bins. In Figure \ref{pic:hr} we provide the HRs measured from the median of the posterior draws with the quoted uncertainties derived from the 17th and 83rd percentiles of the posteriors.

To approximate the absorbing gas column density we use the Portable, Interactive Multi-Mission Simulator ({\sc pimms}) to simulate a power-law spectrum with $\Gamma = 1.8$ subject to a set of column densities ${\rm log }N_{\rm H} = \{20,22,23,23.5,24,24.5 \}$ at the rest-frame of the target, these spectra are then folded through the response curves of Chandra ACIS-S and XMM-PN. The minimum column of $N_{\rm H} = 10^{20}$~cm$^{-2}$ is used to represent foreground galactic absorption, which we include for all of the simulations. The typically low number of detected counts for the ERQ sample considered here necessitate the use of a simple absorbed power-law model over more complex models such as those that can be constructed using {\sc mytorus} \citep{yaqo12}. Using {\sc pimms} we track the expected count-rate as a function of redshift, and use this to predict HR$_1$ and HR$_2$ for Chandra and XMM observations. A particular advantage to harnessing both HR$_1$ and HR$_2$ is that they are sensitive to different levels of incident absorption. HR$_1$ is more sensitive to columns below a few $\times 10^{23}$~cm$^{-2}$, while HR$_2$ is more reliable towards the identification of heavily obscured and Compton-thick columns in the range 0.3--5$\times 10^{24}$~cm$^{-2}$. 

In Figure~\ref{pic:hr}, we provide our BEHR-produced HR$_1$ and HR$_2$ measurements for our ERQ sample as a function of redshift and theoretical tracks of predicted constant $N_{\rm H}$ as a function of redshift produced from our modeling with {\sc pimms} for both the XMM EPIC-PN and Chandra ACIS-S detectors. The significant offset between the predicted HR$_1$ values for PN and ACIS-S at low column densities is due to the lower quantum efficiency at low energies of the Chandra ACIS detectors associated with the deposition of contaminating materials onto the detectors. This contamination is a function of mission time, and given that the majority of the Chandra observations were performed in Cycle 16, these observations are relatively insensitive to photons with $E<1$~keV, significantly reducing the dynamic range of HR$_1$ for the Chandra observations. For the one source which fell on a PN chip-gap, we used the counts observed in MOS1/2, and for consistency with the other ERQs observed in XMM-PN, we applied a model-dependent correction of $-$0.25 to HR1 for this source to convert from MOS to PN. No correction is required to convert HR2 between MOS and PN for the observed low HR2.

When comparing the measured HR$_1$ values to those predicted from the models, we show that all of the detected ERQs are consistent with being moderately- to heavily-obscured, with $N_{\rm H} \gtrsim 3 \times 10^{22}$~cm$^{-2}$. Even at $z \sim 3$, HR$_1$ is insensitive to changes in the observed spectral energy distribution caused by high column densities. Hence, for the five ERQs with HR$_1$ measurements consistent with $N_{\rm H} \gtrsim 3 \times 10^{23}$~cm$^{-2}$, we can harness additional information from the HR$_2$, which is capable of observing steep rises in the SED above rest-frame $\sim$10~keV that is typical of Compton-thick column densities. Indeed, from the combined $N_{\rm H}$ predictions from HR$_1$ and HR$_2$, we find evidence that four of these five ERQs are potentially Compton-thick with column densities of $N_{\rm H} \approx 10^{24}$~cm$^{-2}$. Hence, based on the population of ERQs studied here, we find evidence that ERQs presenting strong outflows are also accompanied by significant gas column densities obscuring their central nuclei. Owing mainly to the relatively large 1$\sigma$ uncertainties (typically $\sim$0.8~dex), the $N_{\rm H}$ values predicted by HR1 and HR2 tend to be consistent. However, we note that there are four sources that are estimated to be strongly obscured with $N_{\rm H} \sim 10^{24}$~cm$^{-2}$ based on HR2, but are predicted to have significantly lower $N_{\rm H}$ based on HR1, and are each inconsistent at the 1--2$\sigma$ level. As we show in the following sections, based on more sophisticated stacking analyses, this could in part be due to an optically-thin scattered component that is contributing to the emission at soft-energies (rest-frame $E<4$~keV) that is not currently being included in our simple absorbed power-law model for converting HR to $N_{\rm H}$.

Given the uncertainty in the intrinsic power-law spectrum and the measurement uncertainty, clearly the column density estimates from the HRs, while highly suggestive of strong obscuration, are uncertain by a factor of a few. We directly demonstrate this uncertainty in the $N_{\rm H}$ estimates in the next section through comparison of the HR predictions and the $N_{\rm H}$ measurements from spectral fits for two of our ERQs. This uncertainty in $N_{\rm H}$ further implies that the intrinsic (obscuration-corrected) luminosities of our sources are very uncertain, as the $2-10$ keV rest-frame luminosity correction depends very sensitively on the value of column density in this regime. For example, as $N_{\rm H}$ ranges from 0.8 to $1.6\times 10^{24}$ cm$^{-2}$, the luminosity correction factor ranges between 14 and 87. With the exception of SDSS~J0006+1215 and SDSS~J2129$-$0018, which have sufficient X-ray counts to perform a spectral fit (Section \ref{sec:ind}), we use the measured HRs to provide an estimate of the absorbing column in each ERQ considered here in order to predict the intrinsic X-ray luminosity at 2--10~keV. As outlined above, the HR$_2$ is not sensitive to $N_{\rm H}$ values below $\approx 10^{23}$~cm$^{-2}$. Hence, for those ERQs that have an HR$_2$ measurement that is consistent (at the 1$\sigma$ level) with no absorption (i.e., HR$_2 \lesssim -0.75$), we choose to use HR$_1$ to estimate $N_{\rm H}$ as these sources likely have lower $N_{\rm H}$ values, otherwise, we use HR$_2$. In Table 2 we provide the predicted $N_{\rm H}$ values for the ERQ sample along with the adopted indicator (i.e., X-ray spectra; HR$_1$; HR$_2$). We use these $N_{\rm H}$ estimates to predict intrinsic $L_{\rm X}$ with their associated associates computed by combining in quadrature the 1$\sigma$ X-ray flux and HR uncertainties.

\subsection{Comments on individual sources}
\label{sec:ind}

Before looking into the average X-ray properties of our sources in Section \ref{sec:stack}, here we discuss sources that are apparent outliers within our sample. Specifically, SDSS~J0006+1215 and SDSS~J2129$-$0018 are the two most strongly detected objects with sufficient counts to enable spectral analysis. SDSS~J0834+0159 is the only non-detected source. Finally, SDSS~J1535+0903 is a redshift outlier with a unique optical spectrum.

In Figure \ref{pic:xray} we show the X-ray spectra of two sources, SDSS~J0006+1215 and SDSS~J2129$-$0018, which have $> 100$ detected counts in their XMM data, sufficient to perform a spectral fit. For both sources, we fit a power-law model combined with foreground Galactic absorption determined from \citet{star92} and a rest-frame absorber which is intrinsic to the source. The unbinned data from the PN and MOS detectors are fit simultaneously using \citet{cash79} statistics, which allows spectral parameter estimation using the maximum likelihood method especially well suited for low count rates. For SDSS~J0006+1215 and SDSS~J2129$-$0018, we find that an absorbed power-law produces a reasonable fit with C-statistic $\sim 181.5$ and 134.9 for 208 and 159 degrees of freedom, respectively. Hence, based on the available data there is no statistical evidence for requiring a more complex model. The best-fit model parameters for SDSS~J0006+1215 are $\Gamma \sim 1.74 \pm 0.07$ and $N_{\rm H} \sim (5.8 \pm 0.7) \times 10^{22}$~cm$^{-2}$. This is consistent with the value of $N_{\rm H}$ of $\sim (6 \pm 2) \times 10^{22}$~cm$^{-2}$ predicted from using HR$_1$. However, the HR$_2$ measurement would suggest a value of $N_{\rm H}$ that is a factor $\sim 3-5$ greater than that measured from the spectroscopy. For SDSS~J2129$-$0018, the best-fit spectral model yields a similar absorbing column of $N_{\rm H} \sim (6.1 \pm 0.6) \times 10^{22}$~cm$^{-2}$, again with a marginally flatter spectral slope of $\Gamma \sim 1.7 \pm 0.1$ than the one we had assumed for the HRs. In this case, the spectroscopic $N_{\rm H}$ value is consistent with the one resulting from our HR analysis. Based on these two sources, we find that the uncertainty in the HR method is large, and $N_{\rm H}$ estimates from the HRs should only be taken as suggestive, not conclusive (especially given the simplicity of the spectral model used in hardness ratio simulations).

One source SDSS~J0834+0159 (with a high-velocity [OIII] outflow discussed by \citealt{zaka16b}) is not detected in any of the X-ray bands. Non-detections can arise because the source is intrinsically X-ray weak, because it is obscured or both. However, we cannot distinguish between these scenarios on the basis of the existing data. 

SDSS~J1535+0903 is at a substantially lower redshift ($z \sim 1.5$) compared with $z\sim 2.3$--3.2 characteristic of the remainder of the ERQs considered here. This object was recognized as having an extremely unusual optical spectrum \citep{ross15} which was analyzed by \citet{wang16b}. The optical spectrum is dominated by high equivalent width Fe II emission which \citet{wang16b} interpret as being resonantly scattered by extended outflows that are viewed nearly edge-on. The [OIII] profile \citep{perr17} shows a strong blue-shifted outflow component. This source has weak apparent X-ray emission and is only marginally detected (S/N$\sim 2$) at $E \sim 1$--7~keV, which, as in the case of  SDSS~J0834+0159, is suggestive of intrinsic X-ray weakness and/or heavy obscuration.

\subsection{Stacking the weakly detected sources}
\label{sec:stack}

\begin{figure}
\includegraphics[width=\linewidth]{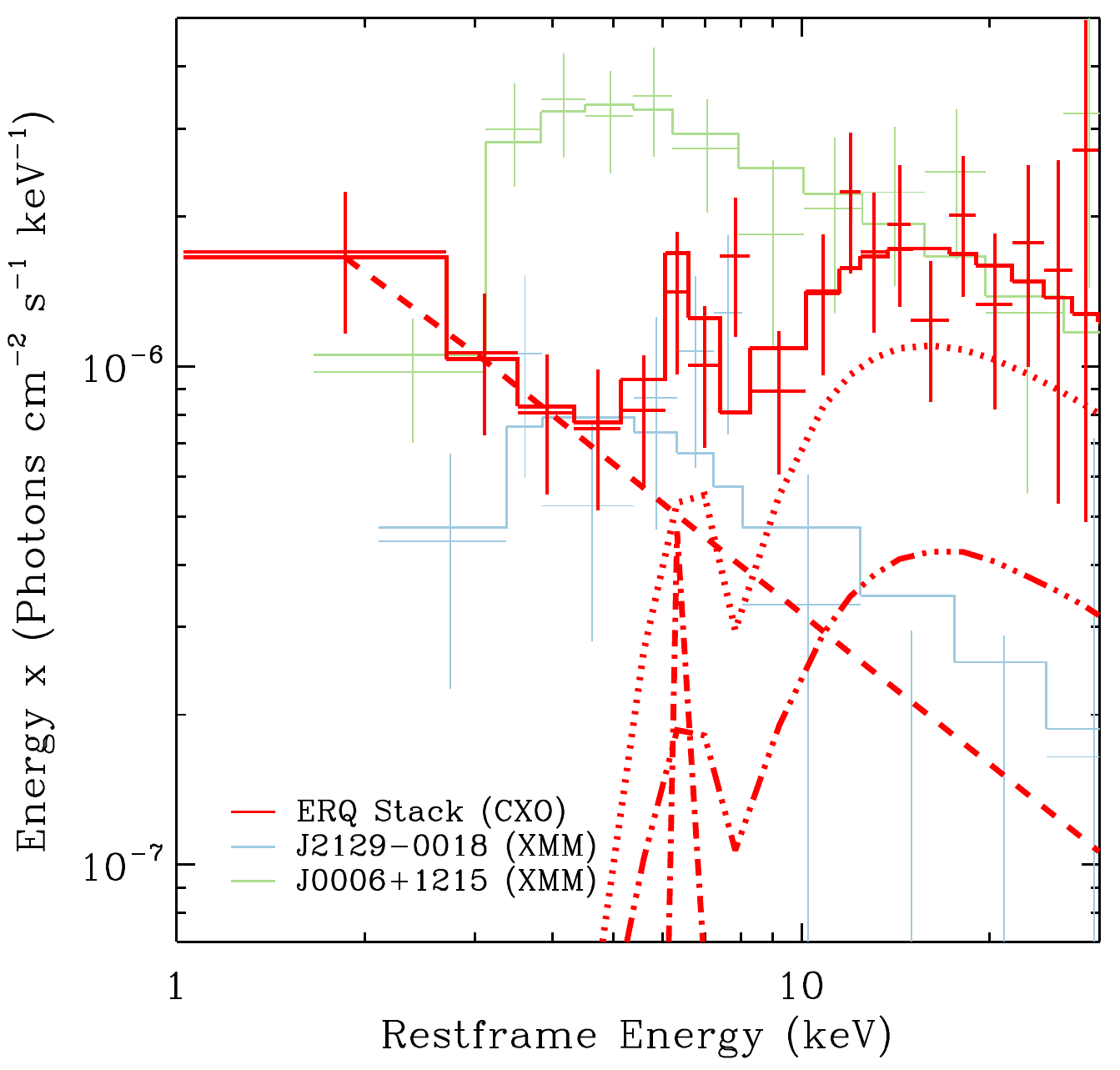}
\caption{Stacked X-ray spectrum of the ERQ and ERQ-like objects that were observed with Chandra (red crosses). The solid red histogram shows the best fit {\tt MYTorus} model composed of a zeroth-order power-law continuum (dotted red line) that is additionally subject to Compton-scattering (dot-dash-dot red line), fluorescent line emission (dash-dot red), and an optically-thin scattered continuum (dashed red line). For comparison purposes, the rest-frame XMM-PN X-ray spectra of the two sources shown in Figure \ref{pic:xray} are provided in light-blue and light-green.}
\label{pic:stack}
\end{figure}

All seven ERQs that were observed with Chandra are only weakly detected in their Chandra X-ray observations, and their low count rates preclude us from analyzing their spectra individually. Instead, we stack these seven sources to determine their average X-ray properties. We use the SDSS spectroscopic redshifts for each of the ERQs to reproject the source aperture photons to the rest-frame energies of the individual sources. We further reproject the individual energy redistribution matrices to the source rest-frames and interpolate onto a common binning. The source aperture photons are then summed across the observations before being combined with the response matrices within {\tt xspec}. We do not include the ERQs covered by XMM observations into the stacked spectrum given the large background differences between XMM and Chandra and the strongly differing responses between the instruments. Furthermore, the inclusion of the two ERQs that are strongly detected in XMM would dominate the stacked spectrum, whereas the purpose of stacking is to determine the average properties of the weakly detected population. The stacked X-ray spectrum contains 138 counts, and is shown in Figure \ref{pic:stack}.

We begin by modeling the stacked X-ray spectrum using a simple power-law subject to local galactic extinction (assumed to be $N_{\rm H} = 10^{20}$~cm$^{-2}$), finding an extremely flat spectral slope of $\Gamma \sim 0.7 \pm 0.2$ with $C_{\rm stat} = 78.8$ and 118 d.o.f. Such flat apparent slopes indicate that the observed spectrum is much harder than the intrinsic power-law spectra of unobscured active nuclei \citep{nand94} and are commonly taken as indicators of absorption \citep{alex01}. Fixing the slope to a more typical value of $\Gamma=1.9$, we use the stacked spectrum to find evidence for intrinsic X-ray absorption of $N_{\rm H} \sim 1.1 \pm 0.2 \times 10^{23}$~cm$^{-2}$, albeit with a slightly reduced significance of $C_{\rm stat} = 88.1$ and 117 d.o.f. The stacked spectrum in Figure \ref{pic:stack} displays a clear inversion of the X-ray spectrum around rest-frame 4~keV, with a rising continuum out to $E \sim 20$~keV, which is also a signature of absorption. Taken together these spectral measurements and features provide evidence that the stacked sources are drawn from a moderately to heavily-obscured quasar population. 

As the data appear to be consistent with significant absorption towards the quasars in our sample, we fit the stacked spectrum with a more physically motivated model using the pre-calculated {\tt MYTorus} tables \citep{yaqo12}, which simulate the reprocessed X-ray emission from a toroidal structure in the heavily obscured and Compton-thick regime. Specifically, we use a {\tt MYTorus} model consisting of a transmitted zeroth-order power-law continuum, a Compton-scattered continuum and emission-line fluorescence due to FeK at $E \sim 6.4$~keV. The advantage of this model is that it self-consistently includes the Compton-scattered component for heavily absorbed sources \citep{lans14}. In the model, we assume an inclination angle of the torus of 85$^{\circ}$, a power-law slope and its normalization, and the column densities for all three components are tied together. The relative normalizations of the Compton-scattered and line-emission components are additionally tied together. Based on our HR--$N_{\rm H}$ analysis in the previous section, there are several weakly detected sources that have apparently contradicting $N_{\rm H}$ estimates from HR1 and HR2 due to significant excess emission at soft-energies. This soft-excess is also observed in the stacked spectrum at $E<4$~keV, and could arise due to a scattered continuum, which has been previously observed in similar populations of red quasars \citep{lama16, glik17}. To account for the observed soft X-ray emission at $E<4$~keV, we additionally include an optically-thin scattered quasar continuum into our spectral model.

The stacked spectrum is well characterized by our physically-motivated model with a best-fit column density for the stacked spectrum of $N_{\rm H} \sim (9.1\pm 3.8)\times 10^{23}$~cm$^{-2}$ and $C_{\rm stat} = 78.5$ for 113 d.o.f.. Therefore, we find that on average the weakly detected ERQs are significantly obscured in the X-ray, and approaching the Compton-thick threshold ($N_{\rm H}\ga 10^{24}$ cm$^{-2}$). The intrinsic power-law is not particularly well-constrained with $\Gamma \simeq 1.55 \pm 0.49$, which is within the standard range of power-law slopes measured in active nuclei. However, for this particular value of $\Gamma$, we find an unusually high and unconstrained scattering fraction of $\sim 19$\%, which we suggest is unlikely. Fixing the spectral slope to $\Gamma = 1.9$, we find a consistent column density to before with $N_{\rm H} \sim (1.1\pm 0.3)\times 10^{24}$~cm$^{-2}$ but with a more realistic, though still relatively high, scattering fraction of $5.9 \pm 2.5$\%. The absorbed flux (at rest-frame 2-10 keV) is $8\times 10^{-15}$ erg~s$^{-1}$~cm$^{-2}$, and the absorption-corrected flux is $4.1\times 10^{-14}$ erg~s$^{-1}$~cm$^{-2}$. At the median redshift of the ERQs observed with Chandra ($z \sim 2.7$), this corresponds to an absorption-corrected rest-frame luminosity of  $L_{2-10 {\rm keV}} \sim 2.6\times 10^{45}$ erg~s$^{-1}$. 

To validate our stacking analysis, we conduct a simultaneous spectral fit directly to all (Chandra and XMM) data using \citet{cash79} statistics. We extract the photons for each ERQ using the same point-source apertures, and treat them in {\sc xspec} as though they were individual source spectra. We then fit the spectra simultaneously with the same combined spectral model, redshifted to the rest-frame of each individual object. We start with a simple absorbed power-law model, finding an extremely flat slope $\Gamma=0.6\pm 0.2$ with an unconstrained column density N$_{\rm H} \sim (0.05 \pm 1.3) \times 10^{22}$ cm$^{-2}$. This is consistent with that found for a simple absorbed power-law fit to our stacked spectrum. As found previously, such a best-fit model is non-physical, and suggestive of a need for a more complex model involving reflection components. We used the same physically motivated {\tt MYTorus} model as above, finding best-fit values for the spectral slope and column density of $\Gamma \sim 1.45 \pm 0.51$ and N$_{\rm H}= (5.5\pm 2.5)\times 10^{23}$ cm$^{-2}$, which are fully consistent with the fit to the stacked X-ray spectrum presented in Fig~\ref{pic:stack}.

\section{Discussion}
\label{sec:disc}

\subsection{X-ray properties of ERQs}

Because of high obscuration, it is possible that none of the observed fluxes of ERQs provide a reliable measure of the intrinsic bolometric accretion luminosity. For ease of comparison with previous work, here we use $\nu L_{\nu}$ at rest-frame 6\micron\ (obtained by power-law interpolating between the observed WISE fluxes) as a measure of quasar luminosity, listed in Table \ref{tab:all}. While this is the most isotropic measure of accretion luminosity currently available to us, theoretical models of quasar obscuration predict that mid-infrared luminosity is not isotropic and can be strongly suppressed in type 2 objects \citep{pier92b}. At low redshifts, type 2 quasars have noticeably redder infrared colors than type 1s \citep{liu13b}: at a fixed 12\micron\ luminosity, 5\micron\ luminosities are a factor of $2-3$ lower in type 2s than in type 1s. In type 1 quasars, the bolometric luminosity can be estimated by multiplying $\nu L_{\nu}[6\micron]$ by a factor of $\sim 7-9$ \citep{rich06}. In our objects, we use the same estimate, but with a caveat that it could be a lower limit on the actual bolometric luminosity if the 6\micron\ luminosity of ERQs is suppressed by optical depth effects.

\begin{figure}
\includegraphics[width=\linewidth]{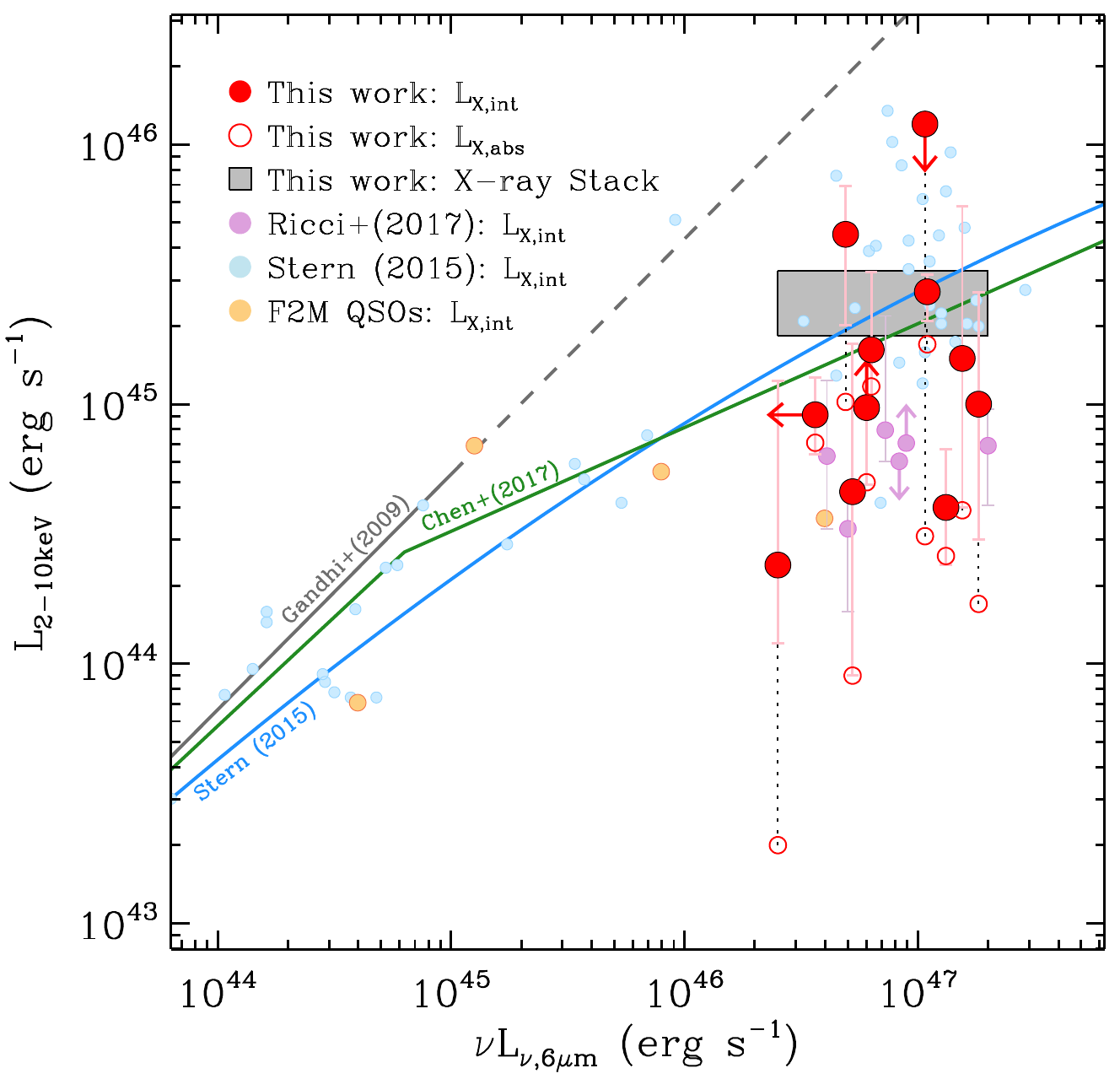}
\caption{The relationship between infrared and intrinsic (absorption corrected) 2--10 keV X-ray luminosities for individual ERQs in this work (red circles). Absorbed (observed) X-ray luminosities for the ERQs are shown with open symbols. Additionally, we show the ERQ X-ray stack (gray shaded region) and further compare with individual type 1 quasars from \citet{ster15}, HotDOGs from \citet{ricc17}, and a sample of lower redshift ($z\sim 0.1$--0.7) reddened FIRST-2MASS quasars presented in \citet{lama16} and \citet{glik17}, shown with blue, purple and orange circles, respectively. The near-linear relation from \citet{gand09} was originally derived for sources with $\nu L_{\nu}[12.3\micron]<10^{45}$ erg s$^{-1}$ and is shown here corrected to 6\micron\ following \citet{chen17} with an extrapolation to higher luminosities shown with a dotted line. Subsequent work demonstrated that the relationship is sub-linear for extremely luminous quasars (blue for \citealt{ster15}, green for \citealt{chen17}). The intrinsic (absorption-corrected) X-ray luminosities of individual ERQs and of the stacked detection are consistent with the relationships derived for type 1 quasars.}
\label{pic:irxrays}
\end{figure}

In Figure \ref{pic:irxrays}, we show the infrared luminosity versus both the observed and estimated intrinsic (absorption-corrected) X-ray luminosities of ERQs in comparison with several other studies. As has been demonstrated by multiple studies, type 1 (unobscured) quasars do not follow the near-linear relationship between IR and intrinsic X-ray luminosities characteristic of lower luminosity Seyfert galaxies ($\nu L_{\nu}$[6\micron]$<10^{45}$ erg~s$^{-1}$; \citealt{gand09}). Extremely luminous type 1 quasars with $\nu L_{\nu}[6\micron]=10^{47}$ erg~s$^{-1}$ show no signs of intrinsic X-ray absorption. However, they lie an order of magnitude below the extrapolation of the linear lower luminosity relationship \citep{ster15,mart17}, calling for revised sub-linear IR-to-X-ray relationships by \citet{ster15} and \citet{chen17}, producing lower $L_{\rm X}$ at fixed $L_{6 \mu m}$. Red quasars of comparable infrared luminosities \citep{bane12, feru14, bane15,lama16,glik17,mart17} are consistent with the same lower X-ray-to-infrared ratios of extremely luminous type 1 quasars, once the X-ray luminosities are corrected for the effects of X-ray obscuration. 

While the origin of this sub-linear relationship between X-ray and infrared luminosities is not well understood, one hypothesis is that X-ray-to-infrared ratios are anti-correlated with Eddington ratios \citep{leig07, luss10, jin12} because a massive accretion flow disrupts the X-ray emitting corona, suppressing the observed X-ray emission, resulting in apparently X-ray weak quasars. At lower luminosities, \citet{will04} also find that near-Eddington sources (narrow-line Seyfert 1 galaxies) have an appreciable X-ray weak sub-population. In the recent NuSTAR serendipitous source survey, the source with the lowest X-ray-to-infrared ratio was a narrow-line Seyfert 1 object \citep{lans17}.

The observed (absorbed) X-ray luminosities of ERQs lie well below the type 1 relationship, with the median object being a factor of $\sim 5$ underluminous in X-rays as compared to the \citet{ster15} and \citet{chen17} relationships. However, at the high column densities found in ERQs the apparent luminosities are strongly affected by intervening absorption. As discussed above, the absorption correction at 2$-$10 keV energies is very sensitive to the assumed column, which is not well constrained for most of our sources. Therefore, our absorption-corrected luminosities are reported in Figure \ref{pic:irxrays} with large uncertainties (which may in fact be underestimated since all of our calculations assume the same intrinsic spectral slope $\Gamma=1.8$). Keeping in mind that our absorption-corrected X-ray luminosities should be considered estimates, we find that the median intrinsic absorption-corrected X-ray luminosity for our sample is $L_{\rm 2-10 keV, int}=10^{45}$ erg~s$^{-1}$, which is on average still a factor of $\sim 2$ below the \citet{ster15} and \citet{chen17} relationships, but consistent with them given the individual uncertainties. 

Furthermore, in Figure \ref{pic:irxrays} we show the absorption-corrected X-ray luminosity of the stacked spectrum, where we have a better-constrained spectral fit, a better $N_{\rm H}$ estimate than in individual sources, and therefore a lower uncertainty on the absorption-corrected luminosity. We find that the intrinsic X-ray luminosity of the stack is almost perfectly in line with the $L_{\rm X}$--$L_{\rm IR}$ relations from \citet{chen17} an \citet{ster15}. This demonstrates that the apparent luminosities of ERQs are likely suppressed by absorption, as opposed to intrinsic weakness in X-rays. We conclude that in the sample of ERQs observed so far there is no evidence for an appreciable population of X-ray weak ERQs. 

Two effects might bias our measurements. One is that, as discussed above, the 6\micron\ luminosity can be suppressed by obscuration. Since we do not correct for this absorption, the apparent 6\micron\ luminosities of our sources could be biased low, with higher intrinsic luminosities. A correction for this effect would move ERQ points and their stack in Figure \ref{pic:irxrays} to the right, making them deficient in X-rays. However, because the $L_{\rm IR}$--$L_{\rm X}$ relationship is rather flat at these luminosities, that correction would have to be large -- as much as an order of magnitude -- to result in an appreciable X-ray deficit. In section \ref{sec:type2} we argue that the hot dust emission in ERQs is not as strongly affected by absorption as in type 2 quasars, so a large absorption correction in the mid-infrared is unnecessary, and even in type 2 quasars this correction is likely small \citep{mateos15,lans17}. Another possible bias may arise in X-ray luminosities. Although we do incorporate the off-axis detections in the serendipitous XMM point-source catalog and in archival {\it Chandra} observations, we do not consider upper limits on X-ray fluxes from off-axis XMM observations (Section \ref{sec:data}). Therefore, our sample may be still marginally biased toward higher X-ray luminosities, and we could be missing a few sources with upper limits on their X-rays in Figure \ref{pic:irxrays}. 

\subsection{Comparison with HotDOGs}
\label{sec:hotdogs}

In terms of the spectral energy distributions (SEDs), luminosities and redshifts, ERQs are similar to hot dust-obscured galaxies (HotDOGs; \citealt{eise12, tsai15, asse15}) -- indeed, the ERQ sample has overlap with HotDOGs targeted for follow-up X-ray observations (Section \ref{sec:data}). The minor differences in their SEDs reflect selection effects: unlike HotDOGs which are selected based on WISE colors alone (see Eisenhardt et al. 2012), ERQs must be bright enough ($r<21.85$ mag) and must have appropriate colors for follow-up spectroscopy in BOSS, as well as be spectroscopically confirmed optical quasars. Therefore, the optical continuum of ERQs is typically stronger than that of HotDOGs, likely because ERQs have a more favorable geometry for scattering, which dominates the observed optical continuum \citep{alex17}. 

Follow-up observations have revealed physical similarities between these two populations. Near-infrared spectroscopy of ERQs reveals high equivalent widths and blueshifts of the [OIII]$\lambda$5007\AA\ line \citep{zaka16b, perr17}, indicating a prevalence of strong outflows of ionized gas. While little equivalent data are available for HotDOGs, one published [OIII]$\lambda$5007\AA\ spectrum \citep{wu17} also shows velocity dispersion above that seen in the local luminous quasar population \citep{zaka14}. With bolometric luminosities well in excess of $10^{47}$ erg~s$^{-1}$ \citep{tsai15, zaka16b}, and estimated black hole masses of $\sim 10^9M_{\odot}$ \citep{wu17}, ERQs and HotDOGs are likely near-Eddington, obscured quasars powered by some of the most massive black holes at $z \sim 3$, and are capable of launching extended ionized gas outflows.  

X-ray properties of several HotDOGs were studied by \citet{ster14a}, \citet{asse16}, and \citet{ricc17}, and \citet{vito18} recently produced a systematic X-ray study and analysis of a significant sample of HotDOGs. These authors find that HotDOGs are a heavily obscured population with a deficiency in their absorption-corrected X-ray luminosities compared to type 1 quasars of similar bolometric luminosity. Another HotDOG presents a high equivalent width Fe K$\alpha$ line and shows other signs of Compton-thick obscuration \citep{pico15}. The population of HotDOGs display X-ray-derived obscuration values extremely similar to those of ERQs. 

X-ray luminosities of HotDOGs from \citet{ricc17} are shown in Figure \ref{pic:irxrays} for comparison with ERQs and other populations. These luminosities have been corrected for absorption using the best available spectral information for individual sources, but despite this correction HotDOGs lie under the best-fit relationships for unobscured luminous sources, with a median absorption-corrected luminosity being a factor of $\sim 3$ below the type 1 relationships. This finding is in contrast to what we see in the ERQ population, which (at the same 6\micron\ luminosity) appears to have X-ray luminosities consistent with the type 1 relationships. \citet{ricc17} suggest that HotDOGs (or some fraction of them) could be intrinsically X-ray weak, similarly to several other populations with known strong outflows, as discussed below. The absorption correction at column densities of $N_{\rm H}=10^{24}$ cm$^{-2}$ is highly uncertain, so one of the HotDOGs shown in Figure \ref{pic:irxrays} has an order of magnitude luminosity uncertainty even though it has a high-quality NuSTAR spectrum \citep{ricc17}, and it thus could be appreciably more luminous than shown (with the quoted luminosity of $\log L_{\rm 2-10 keV}$[erg s$^{-1}$]$=44.9^{+0.86}_{-0.14}$). Future observations will allow to determine whether the X-ray differences between ERQs and HotDOGs persist in larger samples and in better-quality data, despite the close similarity of all other observables between these two populations. 

\subsection{Comparison with BAL quasars, type 1 quasars and ULIRGs}
\label{sec:bal}

While it is not yet clear exactly which selection criterion of ERQs -- red color or high line equivalent width -- predominantly correlates with the presence of outflows, ERQs show outflow activity on a wide range of scales. ERQs are twice as likely as type 1 quasars to show broad absorption-line (BAL) signatures and a hundred times more likely to show $\ge 2500$ km~s$^{-1}$ blueshifts in their CIV$\lambda$1550\AA\ emission relative to the estimated systemic redshift \citep{hama17}. Therefore, in addition to signatures of [OIII]$\lambda$5007\AA\ outflows, which are likely extended on scales of hundreds of parsecs or more, ERQs show a high prevalence of circumnuclear outflows. In this section we compare the X-ray properties of ERQs with those of other quasars with known nuclear and large-scale outflows. 

Initial studies of X-ray emission from BAL quasars found that these objects were highly absorbed in X-rays (${\rm N}_{\rm H}\ga 10^{23}$ cm$^{-2}$) and concluded that this absorption was likely the reason BAL quasars were under-represented in soft X-ray surveys \citep{gall99, gree01, gall02, gall06}. Recent studies have confirmed that BAL quasars are faint X-ray sources, but NuSTAR observations have not found a significant population of Compton-thick BAL quasars, suggesting that a third of BAL quasars are intrinsically X-ray weak \citep{luo13, luo14}, much weaker than non-BAL type 1 quasars of the same optical luminosity. Optical emission-line diagnostics also indicate that in BALs the ionizing emission is softer than an average quasar spectrum \citep{rich11b, bask13, bask15}. These observation are consistent with the theoretical paradigm that X-rays can over-ionize circumnuclear gas, lowering its opacity and suppressing production of radiatively driven winds \citep{murr95, prog00, prog04, sim10}. 

Type 1 quasars with outflows seen in emission lines also tend to have weak X-rays \citep{leig07}. Quasars with the strongest outflow signatures in their CIV tend to have relatively weak emission lines \citep{rich11b}. In a population of such weak-lined quasars examined by \citet{luo15}, half of the objects had X-ray luminosities over an order of magnitude below the \citet{chen17} IR vs X-ray relationship. A similar trend is seen in local ultraluminous infrared galaxies (ULIRGs). Although not all ULIRGs are powered by a luminous active nucleus \citep{sand96}, those that are tend to be highly obscured in X-rays \citep{nard11}. Intriguingly, ULIRGs that show strong signatures of radiatively-driven outflows have X-ray luminosities that are up to two orders of magnitude lower than expected from their bolometric luminosity \citep{teng14, teng15}. 

Thus, the theoretical paradigm in which radiatively driven outflows can only arise when X-rays are relatively weak finds confirmation in known quasars with outflows, as 30-50\% of them are significantly weaker in X-rays than they should be based on relationships shown in Figure \ref{pic:irxrays}. It is therefore puzzling that ERQs -- quasars with known powerful high-velocity ionized gas outflows -- do not appear to follow the same trend. Instead, we find that when their luminosities are corrected for obscuration, they seem to be in line with these relationships. 

Our X-ray observations of ERQs raise the possibility that ERQs and other quasars (non-ERQs) with outflows are not drawn from the same populations. Hence, ERQs may not simply be the obscured analogs of BAL quasars or weak-line type 1 quasars. An alternative explanation is that ERQ winds are not driven by the same mechanism as the line-driven winds which require low X-rays -- for example, ERQ winds could instead be driven by radiation pressure on dust \citep{keat12,thom15}, a phenomenon which \citet{ishi17} suggest may be associated specifically with red quasars. Another possibility is that the combination of high X-ray luminosities and strong outflow activity is due to orientation effects and we would see the weak X-ray population along some other directions (which are also the directions along which gas can be accelerated radiatively), but such objects would not be selected by the same color and magnitude cuts as ERQs. Finally, it is possible that the statistics of the current sample are just too limited and that X-ray-weak ERQs are yet to be discovered. 

\subsection{Comparison with type 2 quasars}
\label{sec:type2}

The color selection that unveiled the ERQ population \citep{ross15} was initially designed to identify high-redshift analogs of the low-redshift obscured quasar populations. In this section we compare X-ray properties of ERQs with those of low-redshift type 2 quasars i.e., optically selected quasars that have no broad component in their H$\alpha$ emission line \citep{zaka03, reye08}. The lack of broad H$\alpha$ in these systems suggests optical extinction over $>10$ mag, sufficient to block all direct light from the broad-line region at the H$\alpha$ wavelength. In contrast, ERQs routinely show broad H$\alpha$ components \citep{zaka16b, wu17}, limiting the amount of extinction toward the broad-line region. This difference between H$\alpha$ properties of type 2 quasars and ERQs can be due to a difference in column densities (larger in type 2s and lower in ERQs) or relative spatial scales of obscuration (larger than the broad-line region in type 2s, vs. smaller than the broad-line region in ERQs). Viewing angle could also produce this difference (closer to edge-on in type 2s than in ERQs), but that should also result in the differences in the intervening X-ray column density.  

Additionally, the SEDs of ERQs rise steeply at a few \micron\ \citep{hama17} and may peak at $5-10$\micron\ \citep{tsai15}. Dust which is emitting at a few \micron\ is confined to scales of a few pc if it is in thermal equilibrium with the quasar radiation \citep{barv87}. In contrast, type 2 quasar SEDs are well represented by smoothly rising power-laws from 3\micron\ to 20\micron, peaking at yet longer wavelengths \citep{mate13, hick17}, suggesting that the hot-dust-emitting region is obscured by material which is optically thick in the  mid-infrared, so that this emission is reprocessed to longer wavelengths. Again, these observations can be explained by a higher level of obscuration in type 2s than in ERQs (e.g., due to a difference in viewing angles). Alternatively, the X-ray absorber could be more compact than the hot-dust emitting region in ERQs and relatively more extended in type 2s. 

X-ray observations allow us to test these hypotheses for the differences between optical and infrared properties of type 2 quasars and ERQs. In an excellent agreement with the geometric unification model \citep{anto93}, low-redshift optically-selected type 2 quasars have high levels of obscuration. \citet{jia13} study a large sample of type 2 quasars at $z<1$ with Chandra and XMM and find that among the 3/4 of the population detected in the X-rays, the median column density is $N_{\rm H}=10^{22.9}$ cm$^{-2}$. The non-detected sources are likely to be Compton-thick rather than X-ray weak, as seen in a few objects with NuSTAR observations \citep{lans15}, with the overall Compton-thick fraction in the type 2 population estimated between 36\% and 76\% \citep{jia13, lans15}. Thus X-ray absorption properties are indistinguishable between type 2 quasars and ERQs. 

The broad-line region and the hot dust region are less obscured in ERQs than they are in type 2 quasars, yet the X-ray-absorbing column densities are similar in these two populations. This similarity rules out the dominant role of viewing angle effects: to make the broad-line region and the hot dust region more visible in ERQs than in type 2, the intervening column density detected in X-rays must be smaller in ERQs, which is not what we see in this population. We conclude that X-ray absorption is confined on scales smaller than a few pc in ERQs, whereas it must be more extended in type 2 quasars. In Section \ref{sec:phys} we discuss the astrophysical implications of this observation. 

Type 2 quasars at $z<1$ studied by \citep{jia13} have lower X-ray luminosities than ERQs (median $\sim 10^{43}$ erg s$^{-1}$ for detected sources), but they are also less bolometrically luminous than ERQs. So instead of directly comparing the X-ray luminosities of these two populations we compare each of them to the type 1 IR-to-X-ray relationships \citep{ster15,chen17} at their respective luminosities. Both populations are hard to detect in the X-rays, but in both cases stacking of weak X-ray sources reveals that their X-ray faintness is due to obscuration and not to intrinsic X-ray weakness. Both $z<1$ type 2 quasars and ERQs are consistent with the luminosity-dependent IR-to-X-ray relationships. 

\subsection{Physical properties of ERQ winds}
\label{sec:phys}

Spectropolarimetric observations of ERQs indicate that essentially all of the observed rest-frame ultra-violet continuum is due to scattered light, as suggested by the high continuum polarization at these wavelengths \citep{alex17}. Furthermore, the kinematic structure of the polarization of ultra-violet emission lines is consistent with scattering produced on scales $\sim 10$~pc in quasi-equatorial winds with large covering factors \citep{veil16, alex17, zaka17} moving with several thousand km~s$^{-1}$. 

The spectral energy distribution of ERQs steeply rises at $1-5$\micron\ \citep{hama17}, indicating a dominant role of emission from hot dust ($T\simeq 500$ K, resulting in a black-body spectrum peak at 6\micron). Taking a $10^{47}$ erg~s$^{-1}$ ionizing luminosity, we find that these temperatures are established at $\sim 12\times (L/10^{47}{\rm erg~s}^{-1})^{1/2}(T/500{\rm K})^{-2}$ pc from the quasar, i.e., on scales that are similar to those of the scattering wind. The difference in the shape of the infrared spectral energy distribution between ERQs (steeply rising to 5\micron\ and then flat) and type 2 quasars (continuously and slowly rising to 30 \micron) suggests that the hot dust emission region cannot be strongly affected by obscuration -- otherwise we would have seen the smoothing of the wavelength dependence of the SED due to radiative transfer effects. Therefore, the high X-ray column must also accumulate on similar scales. 

A natural geometry suggested by these observations is that in which the observed X-ray absorption occurs in the same wind as that seen in spectropolarimetric observations. Compton-thick absorption due to a circumnuclear wind is seen for example in Mrk 231 \citep{brai04}, though this interpretation is disputed \citep{teng14}, and in PDS456 \citep{reev09}. A wind with a column density of $N_{\rm H}=10^{24}$ cm$^{2}$, confined to scales $r<20$ pc, with a covering factor of $\Omega=0.2$ and a velocity of $v=2000$ km~s$^{-1}$ \citep{alex17} has a mass outflow rate 
\begin{eqnarray}
\dot{M}=58M_{\odot}/{\rm year}\times\nonumber\\
\left(\frac{r_{\rm in}}{1{\rm pc}}\right)\left(\frac{N_{\rm H}}{10^{24}{\rm cm}^{-2}}\right)\left(\frac{v}{2000{\rm km\,  s^{-1}}}\right)\left(\frac{\Omega}{0.2}\right).
\end{eqnarray}
Here we have assumed that the wind is in a steady-state with a $\propto 1/r^2$ density profile, so that the column density measurement is weighted toward $r_{\rm in}$, its launching distance. Similar mass outflow rates have been inferred for some X-ray absorbing winds \citep{char09, nard15}, though such winds propagate with much higher velocity ($\ga 0.1c$) and are located at much smaller distances ($\sim 100$ gravitational radii).  

The kinetic power of such wind is 
\begin{equation}
\dot{E}_{\rm kin}=\frac{\dot{M}v^2}{2}=7\times 10^{43}{\rm erg\,  s^{-1}},
\end{equation}
which constitutes a relatively small fraction of the bolometric output. Defining the Eddington mass outflow rate to be $\dot{M}_{\rm Edd}=L_{\rm Edd}/0.1c^2$, we find that such outflow is mildly super-Eddington, with 
\begin{equation}
\frac{\dot{M}}{\dot{M}_{\rm Edd}}=2.6\times \left(\frac{M_{\rm BH}}{10^9M_{\odot}}\right)^{-1}.
\end{equation}
The ratio of outflowing to inflowing mass in near-Eddington accretion is poorly known, but initial simulations suggest that the inflow rates may be a factor of a few higher than the outflow rates \citep{volo15}. Thus ERQs might be accreting at $\sim 10$ times the Eddington limit, but with only mildly super-Eddington emerging luminosities, reflecting the low radiative efficiency of super-Eddington accretion \citep{sado14, jian14b}. 

For the same fiducial parameters, the ratio of the momentum flux of the wind to the available momentum of the photons is
\begin{equation}
\frac{\dot{P}_{\rm wind}}{\dot{P}_{\rm rad}}=\frac{\dot{M}vc}{L}=0.22.
\end{equation}
Therefore, enough photons are in principle available for such wind to be radiatively driven, but the available momentum of the photons would need to be converted to the wind very efficiently. In particular, our fiducial values assume a wind covering fraction of $\Omega=0.2$, so for an emitter which is isotropic on scales $<1$ pc all photon momentum over this covering factor would need to be converted to the wind.

\section{Conclusions}
\label{sec:conc}

In this paper we present X-ray observations of eleven extremely red quasars (ERQs) at $z=1.5-3.2$ selected from the SDSS and WISE data based on their high IR-to-optical flux ratios and high equivalent width CIV emission lines \citep{ross15, hama17}. These objects are among the most luminous quasars at the peak epoch of quasar activity, with directly measured $\nu L_{\nu}[6\micron]$ reaching $10^{47}$ erg~s$^{-1}$ and with inferred bolometric luminosities close to $10^{48}$ erg~s$^{-1}$ \citep{hama17}. Because these values are in excess of the Eddington limit for a $M_{\rm BH}=10^9M_{\odot}$ black hole ($1.3\times 10^{47}$ erg~s$^{-1}$), we hypothesize that these objects are close-to-Eddington or super-Eddington accretors and therefore are likely capable of launching powerful outflows as suggested by numerical simulations \citep{sado14, jian14b}. 

Among the most striking properties of ERQs is the routine occurrence of the [OIII]$\lambda$5007\AA\ emission line with extreme outflow signatures -- strong blue-shifted asymmetries and widths reaching unprecedented FWHM$>5000$ km~s$^{-1}$ \citep{zaka16b, perr17}. This emission line can only originate in regions of relatively low density ($\ll 10^6$ cm$^{-3}$) well outside of the region of gravitational influence of the black hole (likely on scales of hundreds of pc) and is therefore evidence of feedback of quasar activity onto the host galaxy. ERQs also show outflow activity on scales of a few to a few tens of pc, as evidenced by the absorption signatures in CIV$\lambda$1550\AA\ and by the shape of the CIV emission line \citep{hama17}, as well as by spectropolarimetric observations of the rest-frame ultra-violet emission lines \citep{alex17}. While it is not yet known why the combination of the ERQ selection criteria is so successful in identifying objects with strong [OIII] outflows, it may not be surprising because our photometric selection cuts yield objects with extremely high luminosities \citep{hama17} and the high CIV equivalent width requirement may preferentially select objects with outflows \citep{alex17}.

X-ray observations presented here enable us to further test these ideas. Because of the penetrating power of X-rays, we can use X-ray observations to estimate the amount of intervening absorption. Only two objects are detected with sufficient counts to enable spectral fitting; both are absorbed with best-fit $N_{\rm H}\sim 6\times 10^{22}$ cm$^{-2}$. We further use hardness-ratio analyses to find that ERQs are in general a strongly X-ray absorbed population, with estimated column densities in the range $\sim (0.5$--$24) \times 10^{23}$ cm$^{-2}$ and an implied $\sim 50$\% Compton-thick fraction, where we count as potential Compton-thick candidates the five sources with the hardest HR$_2$ hardness ratios and estimated column densities at or above $10^{24}$ cm$^{-2}$. The stack of Chandra observations reveals a spectrum with a best-fit column of $N_{\rm H}=8\times 10^{23}$ cm$^{-2}$, which also supports high levels of average obscuration in the ERQ sample. Finding high levels of obscuration is not surprising, as signs of obscuration are present at other wavelengths and are encoded in the ERQ selection itself. 

Due to the steep rise of the infrared SED of ERQs at a few \micron\ and due to the detection of broad H$\alpha$ components in their near-infrared spectra, we postulate that the observed high X-ray absorption must accumulate on the scales similar to or smaller than the warm-dust emitting region and some of the broad-line region. Further assuming that X-ray absorption is associated with the wind seen in spectropolarimetric observations \citep{alex17} we arrive at estimates for the mass, energy and momentum outflow rates of such wind presented in Section \ref{sec:phys}. We find that the momentum outflow rates are consistent with the availability of photons, but the efficiency of momentum transfer would need to be quite high. While the energy outflow rate is a small fraction of the Eddington limit, the mass outflow rate is marginally super-Eddington, supporting our hypothesis that ERQs are near-Eddington or super-Eddington accretors. 

Another major conclusion of our analysis is that the intrinsic X-ray luminosities of ERQs are largely in agreement with those of type 1 quasars of the same IR power \citep{ster15}. Thus, ERQs do not appear to be intrinsically weak in X-rays, unlike some of the BAL quasars \citep{luo14}, possibly some of the HotDOGs \citep{ricc17}, and some other populations of quasars with known outflows. It has been hypothesized that X-rays may play a critical role in enabling or disabling powerful radiatively-driven outflows \citep{prog00, sim10, rich11b, luo14}: high X-ray luminosities may indicate high accretion rates which are conducive to initiating outflows, but overly high X-ray luminosities may over-ionize the surrounding gas, lower its opacity and suppress outflows. Our observations suggest that X-ray ionization does not appreciably suppress wind activity in ERQs. 

ERQs may represent a somewhat different population from BAL quasars, a third of which show weak intrinsic X-ray luminosities and which are known to be driven via opacity in bound-bound transitions of partially ionized gas. The lack of an appreciable X-ray weak population among ERQs may be a geometric orientation effect, and perhaps the ionizing spectrum seen by the outflowing gas is different from the one seen by the observer. Alternatively, it is possible that the mechanism for wind production is not the same in ERQs and in BAL quasars -- for example, winds in ERQs could be launched by radiation pressure on dust, not gas \citep{keat12, thom15, ishi17}. Observations of a larger sample of ERQs will indicate whether there exists an appreciable population of X-ray weak ERQs. 

\acknowledgments

We thank the anonymous referee for their careful reading of the manuscript, which has helped us to clarify and strengthen several aspects of the analyses. NLZ is grateful to the Institute for Advanced Study for support through the Junior Visiting Professor program and hospitality during multiple subsequent visits which enabled completion of this project, and to Johns Hopkins University for support via the Catalyst Award. Support for this work was also provided in part by the National Aeronautics and Space Administration through Chandra Award Numbers GO5-16107X and GO6-17100X issued by the Chandra X-ray Observatory Center, which is operated by the Smithsonian Astrophysical Observatory for and on behalf of the National Aeronautics Space Administration under contract NAS8-03060. RJA was supported by FONDECYT grant number 1151408. The work of DS was carried out at the Jet Propulsion Laboratory, California Institute of Technology, under a contract with NASA. The scientific results reported in this article are based to a significant degree on observations made by the Chandra X-ray Observatory and data obtained from the Chandra Data Archive.

\bibliography{main.bbl}

\onecolumngrid

\newpage
\begin{sidewaystable}
\begin{center}
\setlength{\tabcolsep}{0.9mm}
\caption{X-ray properties of ERQs, by instrument \label{tab:xray}}
\begin{tabular}{lccccccccccccccccc}
\hline\hline
  \multicolumn{3}{c}{Chandra} &
  \multicolumn{2}{c}{0.3--1 keV} &
  \multicolumn{2}{c}{1--4 keV} &
  \multicolumn{2}{c}{4--7 keV} \\
  \multicolumn{1}{c}{ID} &
  \multicolumn{1}{c}{ObsID} &
  \multicolumn{1}{c}{$t_{\rm GTI}$} &
  \multicolumn{1}{c}{$C_{\rm net}$} &
  \multicolumn{1}{c}{$F_{\rm net}$} &
  \multicolumn{1}{c}{$C_{\rm net}$} &
  \multicolumn{1}{c}{$F_{\rm net}$} &
  \multicolumn{1}{c}{$C_{\rm net}$} &
  \multicolumn{1}{c}{$F_{\rm net}$} &
  \multicolumn{1}{c}{HR$_1^\dagger$} &
  \multicolumn{1}{c}{HR$_2^\dagger$} &
  \multicolumn{1}{c}{$N_{\rm H,HR1}^\ddagger$} &
  \multicolumn{1}{c}{$N_{\rm H,HR2}^\ddagger$} &
  \multicolumn{1}{c}{$L_{\rm X,abs}$} &
  \multicolumn{1}{c}{$L_{\rm X,int}$} &
  \multicolumn{1}{c}{$N_{\rm H}$?$^{\dagger \dagger}$} \\
  \multicolumn{1}{c}{} &
  \multicolumn{1}{c}{} &
  \multicolumn{1}{c}{} &
  \multicolumn{1}{c}{} &
  \multicolumn{1}{c}{($10^{-15}$} &
  \multicolumn{1}{c}{} &
  \multicolumn{1}{c}{($10^{-15}$} &
  \multicolumn{1}{c}{} &
  \multicolumn{1}{c}{($10^{-15}$} &
  \multicolumn{1}{c}{} &
  \multicolumn{1}{c}{} &
  \multicolumn{1}{c}{(log} &
  \multicolumn{1}{c}{(log} &
  \multicolumn{1}{c}{($10^{44}$} &
  \multicolumn{1}{c}{($10^{44}$} &
  \multicolumn{1}{c}{} \\
  \multicolumn{1}{c}{} &
  \multicolumn{1}{c}{} &
  \multicolumn{1}{c}{(ks)} &
  \multicolumn{1}{c}{} &
  \multicolumn{1}{c}{erg/s/cm$^2$)} &
  \multicolumn{1}{c}{} &
  \multicolumn{1}{c}{erg/s/cm$^2$)} &
  \multicolumn{1}{c}{} &
  \multicolumn{1}{c}{erg/s/cm$^2$)} &
  \multicolumn{1}{c}{} &
  \multicolumn{1}{c}{} &
  \multicolumn{1}{c}{cm$^{-2}$)} &
  \multicolumn{1}{c}{cm$^{-2}$)} &
  \multicolumn{1}{c}{erg/s)} &
  \multicolumn{1}{c}{erg/s)} &
  \multicolumn{1}{c}{} \\
\hline
  J0826+0542 & 18206 & 14.86 & $<6.9$        & $<$6.51                & $14.0 \pm 3.7$ & $8.76^{+4.44}_{-3.34}$ & $ 4.9 \pm 2.2$ & $13.0^{+12.5}_{-7.5}$  & 0.88$^{+0.12}_{-0.08}$ & -0.62$^{+0.18}_{-0.29}$ & $23.3^{+1.7}_{-0.2}$ & 23.5$^{+0.3}_{-4.5}$ & 5.0  & 9.7$^{+\infty}_{-4.8}$  & HR1 \\
  J0832+1615 & 18207 & 14.86 & $<6.9$        & $<$7.40                & $ 3.9 \pm 2.0$ & $1.73^{+1.93}_{-1.10}$ & $ <5.4$        & $<10.3$                & 0.80$^{+0.16}_{-0.10}$ & -0.26$^{+0.22}_{-0.23}$ & $23.1^{+0.3}_{-0.4}$ & 24.0$^{+0.2}_{-0.3}$ & 0.9  & 4.6$^{+12.5}_{-3.7}$  & HR2 \\
  J1535+0903 & 18208 & 14.86 & $<5.4$        & $<$5.75                & $ 1.8 \pm 1.4$ & $1.08^{+2.04}_{-0.91}$ & $ <6.9$        & $<13.2$                & 0.47$^{+0.39}_{-0.25}$ & -0.62$^{+0.20}_{-0.38}$ & $21.8^{+1.3}_{-2.8}$ & 23.3$^{+0.4}_{-4.3}$ & 0.2  & 2.4$^{+9.9}_{-1.2}$  & HR1 \\
  J1652+1728 & 18205 & 15.31 & $<5.4$        & $<$6.25                & $ 9.9 \pm 3.1$ & $5.27^{+3.31}_{-2.32}$ & $ <5.4$        & $<10.0$                & 0.57$^{+0.43}_{-0.23}$ & -0.02$^{+0.50}_{-0.46}$ & $22.1^{+2.9}_{-3.1}$ & 23.9$^{+0.4}_{-0.6}$ & 3.9  & 15.0$^{+43}_{-11}$ & HR2 \\
  J0220+0137 & 18708 & 31.32 & $<6.9$        & $<$3.05                & $ 6.8 \pm 2.6$ & $1.96^{+1.57}_{-1.02}$ & $ 2.9 \pm 1.7$ & $3.27^{+4.52}_{-2.36}$ & 0.65$^{+0.29}_{-0.17}$ & -0.16$^{+0.30}_{-0.32}$ & $22.8^{+0.7}_{-0.7}$ & 24.2$^{+0.2}_{-0.4}$ & 1.7  & 10.0$^{+17}_{-7.0}$ & HR2 \\
  J0915+5613 & 04821 & 23.00 & $1.6 \pm 1.4$ & $1.73^{+3.76}_{-1.52}$ & $35.9 \pm 6.0$ & $14.1^{+4.4}_{-3.5}$   & $12.5 \pm 3.6$ & $21.7^{+12.0}_{-9.0}$  & 0.88$^{+0.09}_{-0.06}$ & -0.29$^{+0.14}_{-0.15}$ & $23.3^{+0.3}_{-0.1}$ & 24.0$^{+0.1}_{-0.3}$ & 10.2 & 45.1$^{+24}_{-25}$ & HR2 \\
  J1121+5705 & 06958 & 4.72  & $6.0 \pm 2.4$ & $10.3^{+8.7}_{-5.6}$   & $12.9 \pm 3.6$ & $24.9^{+13.4}_{-9.8}$  & $ 1.9 \pm 1.4$ & $17.6^{+31.7}_{-14.1}$ & 0.52$^{+0.22}_{-0.19}$ & -0.54$^{+0.19}_{-0.26}$ & $22.3^{+0.8}_{-3.3}$ & 23.5$^{+0.3}_{-4.5}$ & 11.7 & 16.2$^{+16}_{-7.0}$ & HR1 \\
\hline\end{tabular}

\hspace{0.5cm}

\begin{tabular}{lccccccrccccrcccccccc}
\hline
  \multicolumn{3}{c}{XMM PN} &
  \multicolumn{2}{c}{0.3--1 keV} &
  \multicolumn{2}{c}{1--4 keV} &
  \multicolumn{2}{c}{4--7 keV} \\
  \multicolumn{1}{c}{ID} &
  \multicolumn{1}{c}{ObsID} &
  \multicolumn{1}{c}{$t_{\rm GTI}$} &
  \multicolumn{1}{c}{$C_{\rm net}$} &
  \multicolumn{1}{c}{$F_{\rm net}$} &
  \multicolumn{1}{c}{$C_{\rm net}$} &
  \multicolumn{1}{c}{$F_{\rm net}$} &
  \multicolumn{1}{c}{$C_{\rm net}$} &
  \multicolumn{1}{c}{$F_{\rm net}$} &
  \multicolumn{1}{c}{HR$_1$} &
  \multicolumn{1}{c}{HR$_2$} &
  \multicolumn{1}{c}{$N_{\rm H,HR1}$} &
  \multicolumn{1}{c}{$N_{\rm H,HR2}$} &
  \multicolumn{1}{c}{$L_{\rm X,abs}$} &
  \multicolumn{1}{c}{$L_{\rm X,int}$} &
  \multicolumn{1}{c}{$N_{\rm H}$?} \\
  \multicolumn{1}{c}{} &
  \multicolumn{1}{c}{} &
  \multicolumn{1}{c}{} &
  \multicolumn{1}{c}{} &
  \multicolumn{1}{c}{($10^{-15}$} &
  \multicolumn{1}{c}{} &
  \multicolumn{1}{c}{($10^{-15}$} &
  \multicolumn{1}{c}{} &
  \multicolumn{1}{c}{($10^{-15}$}  &
  \multicolumn{1}{c}{} &
  \multicolumn{1}{c}{} &
  \multicolumn{1}{c}{(log} &
  \multicolumn{1}{c}{(log} &
  \multicolumn{1}{c}{($10^{44}$} &
  \multicolumn{1}{c}{($10^{44}$} &
  \multicolumn{1}{c}{} \\
  \multicolumn{1}{c}{} &
  \multicolumn{1}{c}{} &
  \multicolumn{1}{c}{(ks)} &
  \multicolumn{1}{c}{} &
  \multicolumn{1}{c}{erg/s/cm$^2$)} &
  \multicolumn{1}{c}{} &
  \multicolumn{1}{c}{erg/s/cm$^2$)} &
  \multicolumn{1}{c}{} &
  \multicolumn{1}{c}{erg/s/cm$^2$)} &
  \multicolumn{1}{c}{} &
  \multicolumn{1}{c}{} &
  \multicolumn{1}{c}{cm$^{-2}$)} &
  \multicolumn{1}{c}{cm$^{-2}$)} &
  \multicolumn{1}{c}{erg/s)} &
  \multicolumn{1}{c}{erg/s)} &
  \multicolumn{1}{c}{} \\
\hline
  J0834+0159 & 762260101 & 22.8     & $<22.4$        & $<1.08$                & $<31.5$          & $<5.35$                & $<22.7$        & $<10.4$              & 0.06$^{+0.72}_{-0.51}$ &  0.02$^{+0.65}_{-0.59}$ & $22.1^{+1.0}_{-3.1}$ & $24.1^{+0.5}_{-0.5}$ & $<3.1$ & $<120$ & HR2  \\
  J1310+3225 & 020540401 & chip-gap &     -          &        -               & -                & -                      &  -             & -                    & 0.61$^{+0.28}_{-0.20}$ & -0.80$^{+0.10}_{-0.20}$ & $23.0^{+0.4}_{-0.3}$ & $20.0^{+1.9}_{-1.0}$ & 2.6 & 4.0$^{+2.7}_{-1.6}$ & HR1 \\
  J2129-0018 & 729160501 & 31.4     & $29.3 \pm 7.6$ & $1.02^{+0.26}_{-0.27}$ & $ 56.3 \pm 11.1$ & $6.94^{+1.37}_{-1.37}$ & $<19.9$        & $<6.59$              & 0.26$^{+0.22}_{-0.20}$ & -0.78$^{+0.10}_{-0.22}$ & $22.6^{+0.3}_{-0.4}$ & $20.0^{+2.1}_{-1.0}$ & 7.1 & 9.1$^{+3.6}_{-2.7}$ & Spec \\
  J0006+1215 & 763780701 & 12.7     & $37.5 \pm 7.9$ & $3.25^{+0.68}_{-0.69}$ & $136.0 \pm 13.6$ & $41.4^{+4.2}_{-4.1}$   & $33.9 \pm 7.5$ & $27.8^{+6.1}_{-6.2}$ & 0.55$^{+0.08}_{-0.07}$ & -0.49$^{+0.09}_{-0.09}$ & $22.7^{+0.1}_{-0.1}$ & 23.5$^{+0.1}_{-0.4}$ & 17.0 & 27.1$^{+4.5}_{-6.2}$ & Spec \\
\hline\end{tabular}

\begin{tabular}{lccccccccccccccc}
\hline
  \multicolumn{3}{c}{XMM MOS1-2} &
  \multicolumn{2}{c}{M1 0.3--1 keV} &
  \multicolumn{2}{c}{M1 1--4 keV} &
  \multicolumn{2}{c}{M1 4--7 keV} &
  \multicolumn{2}{c}{M2 0.3--1 keV} &
  \multicolumn{2}{c}{M2 1--4 keV} &
  \multicolumn{2}{c}{M2 4--7 keV} \\
  \multicolumn{1}{c}{ID} &
  \multicolumn{1}{c}{ObsID} &
  \multicolumn{1}{c}{$t_{\rm GTI}$} &
  \multicolumn{1}{c}{$C_{\rm net}$} &
  \multicolumn{1}{c}{$F_{\rm net}$} &
  \multicolumn{1}{c}{$C_{\rm net}$} &
  \multicolumn{1}{c}{$F_{\rm net}$} &
  \multicolumn{1}{c}{$C_{\rm net}$} &
  \multicolumn{1}{c}{$F_{\rm net}$} &
  \multicolumn{1}{c}{$C_{\rm net}$} &
  \multicolumn{1}{c}{$F_{\rm net}$} &
  \multicolumn{1}{c}{$C_{\rm net}$} &
  \multicolumn{1}{c}{$F_{\rm net}$} &
  \multicolumn{1}{c}{$C_{\rm net}$} &
  \multicolumn{1}{c}{$F_{\rm net}$} \\
  \multicolumn{4}{c}{} &
  \multicolumn{1}{c}{($10^{-15}$} &
  \multicolumn{1}{c}{} &
  \multicolumn{1}{c}{($10^{-15}$} &
  \multicolumn{1}{c}{} &
  \multicolumn{1}{c}{($10^{-15}$} &
  \multicolumn{1}{c}{} &
  \multicolumn{1}{c}{($10^{-15}$} &
  \multicolumn{1}{c}{} &
  \multicolumn{1}{c}{($10^{-15}$} &
  \multicolumn{1}{c}{} &
  \multicolumn{1}{c}{($10^{-15}$}\\
  \multicolumn{2}{c}{} &
  \multicolumn{1}{c}{(ks)} &
  \multicolumn{1}{c}{} &
  \multicolumn{1}{c}{erg/s/cm$^2$)} &
  \multicolumn{1}{c}{} &
  \multicolumn{1}{c}{erg/s/cm$^2$)} &
  \multicolumn{1}{c}{} &
  \multicolumn{1}{c}{erg/s/cm$^2$)} &
  \multicolumn{1}{c}{} &
  \multicolumn{1}{c}{erg/s/cm$^2$)} &
  \multicolumn{1}{c}{} &
  \multicolumn{1}{c}{erg/s/cm$^2$)} &
  \multicolumn{1}{c}{} &
  \multicolumn{1}{c}{erg/s/cm$^2$)} \\

\hline
  J0834+0159 & 762260101 & 27.8 & $<7.4$         & $<1.38$ & $<8.0$       & $<2.91$                & $<4.3$ & $<4.90$ & $<6.1$ & $<1.13$ & $<6.84$      & $<2.47$                & $<3.1$ & $<3.53$ \\ 
  J1310+3225 & 020540401 & 47.9 & $<9.8$         & $<1.06$ & 15.2$\pm$5.4 & $3.21^{+1.13}_{-1.14}$ & $<7.6$ & $<5.02$ & $<8.7$ & $<0.94$ & 14.9$\pm$5.4 & $3.14^{+1.14}_{-1.14}$ & $<8.1$ & $<5.36$ \\ 
  J2129-0018 & 729160501 & 34.3 & 10.9$\pm$4.1 & $1.64^{+0.60}_{-0.62}$ & 26.2$\pm$6.4 & $7.71^{+1.86}_{-1.90}$ & $<9.5$ & $<8.78$ & 11.2$\pm$4.1 & $1.69^{+0.60}_{-0.64}$ & 30.2$\pm$6.4 & $8.89^{+1.80}_{-1.92}$ & 8.2$\pm$3.6 & $7.58^{+3.22}_{-3.33}$ \\ 
  J0006+1215 & 763780701 &  17.2 & 11.4$\pm$4.0 & 3.42$^{+1.20}_{-1.22}$ & 62.4$\pm$8.6 & 38.6$^{+6.1}_{-7.0}$ & 10.2$\pm$3.9 & 18.8$^{+7.2}_{-7.2}$ & 9.6$\pm$3.5 & 2.89$^{+1.02}_{-1.10}$ & 72.4$\pm$8.9 & 42.5$^{+5.2}_{-5.3}$ & 8.3$\pm$3.3 & 15.3$^{+6.1}_{-6.1}$ \\
\hline
\end{tabular}
\end{center}
$^\dagger$ Hardness ratios as defined in Section~\ref{sec:analysis}. $^\ddagger$ Column density inferred from hardness ratio diagnostics. $^{\dagger\dagger}$ The diagnostic (HR1; HR2; X-ray spectra) used to determine the final value of $N_{\rm H}$ adopted in order to calculate the intrinsic X-ray luminosity ($L_{\rm X,int}$).

\end{sidewaystable}

\end{document}